\documentclass[preprint]{aastex}
\usepackage{color, soul}
\usepackage{graphicx}
\usepackage{subfigure}
\usepackage{rotating}
\usepackage{subfigure}
\usepackage{longtable}
\usepackage{lscape}
\usepackage{amssymb}
\usepackage[titletoc]{appendix}
\usepackage{multirow}

\begin{document}
\title{Planetary Nebulae in the Circumnuclear Region of M31: A Spectroscopic Sample}
\author{Anqi Li\altaffilmark{1,2}, Zhiyuan Li\altaffilmark{1,2}, Hui Dong\altaffilmark{3}, Xuan Fang\altaffilmark{4,5}\footnote{*}, Xiao-Jie Xu\altaffilmark{1,2}}
\affil{$^{1}$ School of Astronomy and Space Science, Nanjing University, Nanjing 210046, China; lizy@nju.edu.cn}
\affil{$^{2}$ Key Laboratory of Modern Astronomy and Astrophysics (Nanjing University), Ministry of Education, Nanjing 210046, China}
\affil{$^{3}$ Instituto de Astrof\'{i}sica de Andaluc\'{i}a (CSIC), Glorieta de la Astronom$\acute{a}$  S/N, 18008,Granada, Spain}
\affil{$^{4}$ Laboratory for Space Research, Faculty of Science, University of Hong Kong, Pokfulam Road, Hong Kong, China}
\affil{$^{5}$ Department of Earth Sciences, Faculty of Science, University of Hong Kong, Pokfulam Road, Hong Kong, China}
\footnotetext{* Visiting Astronomer, Key Laboratory of Optical Astronomy, National Astronomical Observatories, Chinese Academy of Sciences (NAOC), 20A Datun Road, Chaoyang District, Beijing 100101, China}
%
%
%

\begin{abstract}
Planetary nebulae (PNe) are an important tool for studying the 
dynamics and chemical evolution of galaxies in the Local Universe, 
given their characteristic, bright emission line spectra.  The 
Andromeda Galaxy (M31) provides a unique laboratory for studying 
PNe in the circumnuclear region, thanks to its proximity and almost 
uniformly low line-of-sight extinction that  ensures observations 
with high resolution and sensitivity. Using the WIYN/Hydra multi-fiber 
spectrograph, we have obtained optical (4119--6882\,{\AA}) spectra 
of 77 PN candidates selected from {\it Hubble Space Telescope} 
narrow-band imaging, which are located within the central 
$\sim$500~pc region of M31.  Among these candidates, 49 (64\%) are 
spectroscopically observed for the first time.  The spectra of 300 
previously known PNe and H~{\sc ii} regions, which primarily reside 
in the disk, are also taken for comparison.  All 77 circumnuclear PN 
candidates exhibit prominent emission lines, including [O~{\sc iii}] 
$\lambda$5007, [N~{\sc ii}] $\lambda$6583, H$\alpha$ and H$\beta$, 
strongly suggesting that they are genuine PNe.  We measured the line 
fluxes, radial velocities and line widths for all objects, and 
found that the radial velocities of the circumnuclear PNe generally 
trace rotation of the inner bulge. We also estimated a dynamical mass 
of $\sim$6.4$\pm$0.5$\times$10$^{9}$\,$M_{\odot}$ enclosed within an 
effective galactocentric radius of 340~pc, which is compatible with 
the previously estimated total stellar mass in this region.
\end{abstract}


\section{Introduction}

Planetary nebulae (PNe) are the end products of the evolution of 
low- and intermediate-mass {\bf ($\sim$1--8\,$M_{\sun}$)} stars.  At 
this stage, the stars eject metal-rich materials into the interstellar 
medium (ISM), which is a fundamental process in the chemical evolution 
of galaxies \citep{Kwok00}.  The stellar ejecta is ionized by 
ultraviolet (UV) photons from the central hot stars, producing 
significant emission lines, in particular [O~{\sc iii}], [N~{\sc ii}]
and H$\alpha$ (\citealt{Kwok00,Osterbrock06}), which provide useful 
diagnostics for the stellar metallicity (e.g., 
\citealp{Delgado-Inglada16}).  When found in galactic spheroids, PNe 
can also serve as a robust tracer of stellar kinematics by using 
their radial velocities derived from the emission lines, and 
consequently a useful probe of the dynamical mass (e.g., 
\citealp{Merrett06a,Coccato09,Herrmann09}). 

To date, PN candidates have been routinely identified in nearby, 
typically early-type, galaxies out to a distance of tens of Mpc, 
often through their bright [O~{\sc iii}] $\lambda$5007 nebular 
emission line (e.g., 
\citealp{Ciardullo04,Coccato09,Sarzi11,Longobardi15a,Longobardi15b}),
which can reprocess up to {$\sim$10\%} of the central star's 
bolometric luminosity \citep{Dopita92,Sch10}. 
However, identification of PNe in galactic nuclei, where strong stellar background is present, has proven a challenging task. 
PN surveys have been conducted to within the innermost $\sim$100 pc of M81 \citep{Jacoby89}, NGC 5128 \citep{Walsh99,Peng04,Walsh15}, M33 \citep{Ciardullo04} and NGC 5102 \citep{McMillan94}, offering insight to the dynamics and metallicity in the unique environment of galactic nuclei. 

Thanks to its proximity (785~kpc, where 1$^{\prime\prime}$ corresponds
to 3.8~pc; \citealt{McConnachie05}), which effectively reduces the 
stellar background within an imaging pixel, and its uniformly low 
line-of-sight extinction ($A_{\rm V} \lesssim 0.5$; \citealp{Dong16}),
M\,31 holds promise for identifying circumnuclear PNe with 
high-resolution, high-sensitivity optical imaging and spectroscopy. 
Several spectroscopic surveys for PNe in M31 have been conducted.
The narrow-band survey of \cite{Ciardullo89} found 429 PNe in the bulge, but follow-up spectroscopic observations \citep{Richer99, Jacoby99} were only focused on the brightest ones.
\cite{Halliday06} presented velocities for a sample of 723 PNe in the disk and bulge, based on the [O III]$\lambda$4959,5007 doublet obtained by the WYFFOS fibre spectrograph with a spectral resolution of 0.9\,\AA. Merrett et al.~(2006; hereafter MMD06) conducted a survey of [O III]$\lambda$5007 emission line objects, using the Planetary Nebula Spectrograph (PN.S) with a spectral resolution of 1.85\,{\AA}, and catalogued 2615 PN candidates. \cite{Sanders12} presented a spectroscopic follow-up of the PNe in MMD06 and \cite{Halliday06}, and derived the metallicity profile along the disk. Their observations were taken by the MMT/Hectospec multi-fiber spectrograph, with a wide wavelength range of 3650--9200\,{\AA} and a modest spectral resolution of 5\,{\AA}.  Using LAMOST spectroscopic data, \cite{Yuan10} discovered tens of PNe in the outskirts of M31. However, the majority of the above objects reside in the disk and outer bulge/halo of M\,31; 
 only a few tens of PNe were found within the central 500 pc. 
On the other hand, using the SAURON integral-field spectroscopic observation, \cite{Pastorello13} identified 12 PN candidates within a projected distance of  21$\arcsec$  ($\sim$80 pc) from the center of M\,31, again through the detection of [O III]$\lambda$5007. This work, however, suffered from a small field-of-view and low spatial resolution, thus only revealing the brightest PNe in this very central region.

Recently, we carried out an \emph{HST}/WFC3 program to map diffuse, 
ionized gas in the circumnuclear region of M31 (\citealt{Dong14,
Dong16}).  Combining the WFC3/F502N and WFC3/F547M images obtained 
in this program, we have identified a 
total of 249 PN candidates located within a projected galactocentric 
radius of $\sim$500~pc, the majority of which are new identifications 
(A.~Li et al.\ 2018, in preparation; hereafter Paper~I).  In this 
work, we present the results of follow-up optical spectroscopy for 
77 PN candidates, using the WIYN/Hydra multi-object spectrograph. 
In Section~2, we describe our sample selection, observations and data
reduction. In Section~3, we analyze the most prominent emission lines
detected in the spectra; these candidates are identified to be 
genuine PNe based on these emission lines, and their individual and 
statistical properties are derived, including line ratios and 
kinematics.  In Section~4, we discuss the implications of this 
sizable sample of circumnuclear PNe.  We draw our conclusions in 
Section~5.  Throughout this work, we quote errors at the 
1$\sigma$ (68.3\%) confidence level, unless otherwise stated. 

\begin{figure*} 
\centering
\subfigure{
\includegraphics[scale=0.6,width=2.75in,angle=0,]{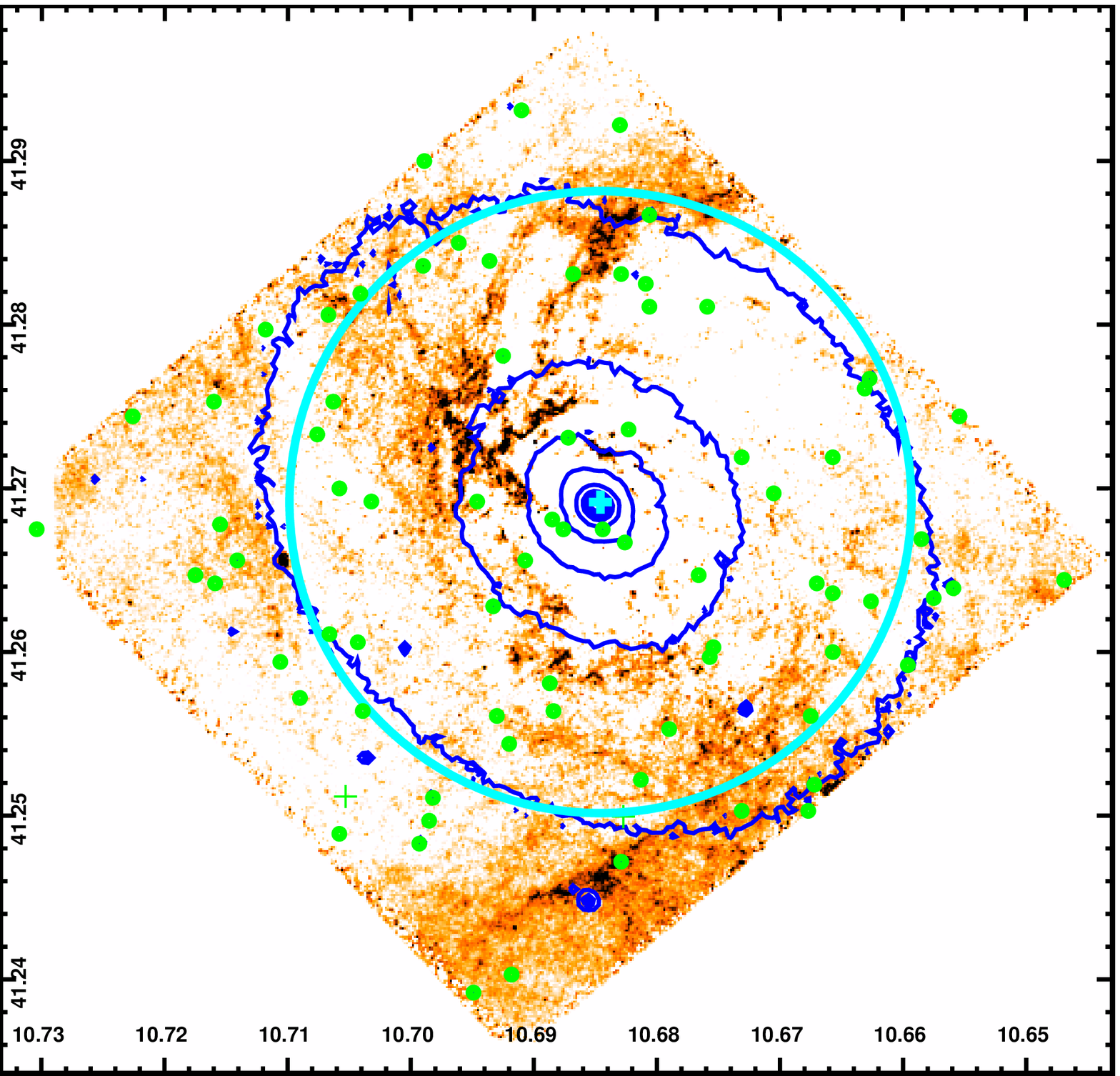}
}
\subfigure{
\includegraphics[width=2.7in,angle=0]{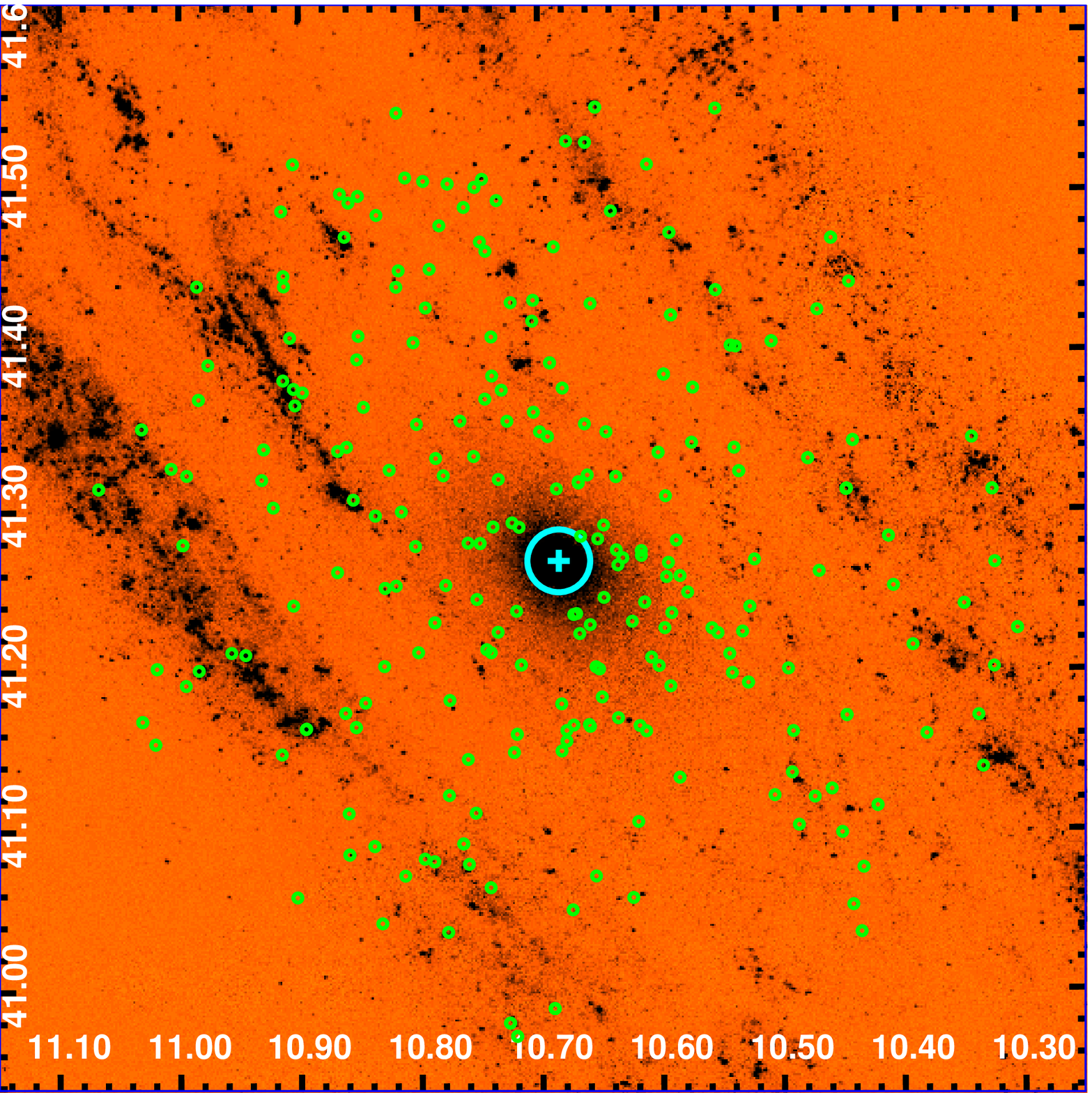}
\quad
}
\caption{Left panel: An intensity map of diffuse H$\alpha$ emission obtained from \emph{HST} WFC3/F656N image (Z. Li et al. in preparation). 77 PN candidates in this paper are marked with green circles. The blue contours outline the intensity distribution of the stellar bulge of M31. Right panel: GALEX far-UV image of M31 \citep{Gil07}.  The 300 MMDOs observed by WIYN/Hydra 
are marked with green circles.  In both panels, darker color means 
higher surface brightness. The center of M31 is marked by a cyan 
cross, and cyan circles in both panels have a radius of 
70$\arcsec$ ($\sim$267~pc). HPNe detected within this radius are 
used for dynamical mass estimate of the bulge 
(Section~\ref{sec:mass}). The X- and Y-axes are RA and DEC, 
in epoch 2000, respectively.}
\label{fig:bulgemap}
\end{figure*}

\section{Observations and Data Preparation} \label{sec:data}

\subsection{Sample selection} \label{subsec:X-ray data}

In Paper I, we have identified 249 PNe candidates in the 
circumnuclear region of M\,31, by contrasting the \emph{HST} 
WFC3/UVIS F502N and F547M images. These candidates, hereafter 
referred to as HPNe, serve as the foremost sample for the 
follow-up spectroscopy. To facilitate a comparison with the PN 
population in the outer regions,
we also added to our target list the 500 brightest objects (sorted 
by the apparent magnitude $m_{\lambda5007}$) in the MMD06 catalog 
(hereafter MMDOs), which primarily consist of PNe, but also contain 
a small number of H~{\sc ii} regions. The galactocentric radii of our 
targets thus range from 10 pc to 10 kpc (Figure~\ref{fig:bulgemap}).

 \subsection{WIYN/Hydra observations} \label{sec:detection}

On 2014 October 3 and 2016 September 30, we observed the targets 
using the multi-object spectrograph, Hydra, on the 3.5m WIYN 
telescope \citep{Barden95} at Kitt Peak. 
 The focal plane of Hydra corresponds to a circle of $1^{\circ}$-diameter on the sky, which, at the time of the observations, could accommodate up to 90 fibers in a single field.
The observations used the red fibers of 2$\arcsec$-diameter, which feed the Bench spectrograph with the 600 lines mm$^{-1}$ grating. The wavelength coverage was 4119\,\AA--6882\,\AA, with a FWHM resolution of 3.35\,{\AA}.

All observed fields were centered at the position of the M\,31 nucleus. 
We used fiber-assignment program {\sl whydra} from WIYN observatory to assign fiber positions. 
Due to the minimally allowed angular separation (37$\arcsec$) between any two fibers, crowding of the HPNe limits the observing efficiency in a single field. 
In each field, most fibers were positioned to cover a PN, and a few fibers were devoted to ``random sky" positions to determine the sky background. 
While this choice limits our ability to accurately determining the immediate background of the HPNe (see further discussion in Section~\ref{sec:line}), this ensures a good observing efficiency.  
Hence we took a total of 8 fields to cover 77 HPNe ($\sim$30\% of our entire candidate list), with priority given to the brighter (in terms of m$_{5007}$) ones. In addition, a total of 300 MMDOs were covered in these fields, among which 267 have been classified as PNe and 33 as HII regions by MMD06. Figure~\ref{fig:bulgemap} illustrates the target positions. 
The exposure time varied between $3\times1200$\,s, $4\times1200$\,s or $3\times1800$\,s, depending on the presumed [O III]$\lambda$5007 magnitude of the targets (Paper I). 
Several spectroscopic standard stars were also observed between the science exposures, for flux calibration. 
A log of the Hydra observations is given in Table~\ref{oblog}.

\begin{deluxetable}{ccccccccccccccc}
\tablecaption{Observation log}
\tablewidth{0pt}
\tablehead{
\colhead{Field} &
\colhead{RA} &
\colhead{DEC} &
\colhead{Date} &
\colhead{Exposure} \\
\colhead{(1)} &
\colhead{(2)} &
\colhead{(3)} &
\colhead{(4)} &
\colhead{(5)} &
 }
\startdata
1& 00:42:47.6&  +41:16:37.4 &2014-10-03&  3$\times$1800\,s\\ \\
2& 00:42:44.7&  +41:16:07.6 &2014-10-03&  4$\times$1200\,s\\ \\
3& 00:42:49.5&  +41:14:36.8 &2014-10-03&  3$\times$1200\,s\\ \\
4& 00:42:44.4&  +41:16:07.7 &2016-09-30&  3$\times$1200\,s\\ \\
5& 00:42:44.1&  +41:16:09.5 &2016-09-30&  3$\times$1800\,s\\ \\
6& 00:42:44.1&  +41:16:08.2 &2016-09-30&  4$\times$1200\,s\\ \\
7& 00:42:44.2&  +41:16:09.1 &2016-09-30&  3$\times$1200\,s\\ \\
8& 00:42:44.4&  +41:16:07.7 &2016-09-30&  3$\times$1200\,s\\ \\
\enddata

\tablecomments{(1): Each field has an independent set of targets. (2)-(3): Field center, in epoch 2000; (4) UT date of observation; (5) Exposure time in units of seconds.}
\label{oblog}
\end{deluxetable}

Data reduction followed the standard procedure involving several 
{\sc iraf}\footnote{{\sc iraf}, the Image Reduction and Analysis 
Facility, is distributed by the National Optical Astronomy 
Observatory, which is operated by the Association of Universities 
for Research in Astronomy under cooperative agreement with the 
National Science Foundation.} tasks.  First, we used the {\sl 
ccdproc} command in the {\sl ccdred} package for image trimming, 
de-bias and dark current corrections.  The data were then combined 
using {\sl imcombine} in the {\sl immatch} package to remove 
cosmic-rays and hot pixels.  The {\sl dohydra} task was then applied 
for spectral extraction, pixel-response correction and wavelength 
calibration.  The {\sl sarith} application in the {\sl onedspec} 
package was then used for fiber-responding calibration.  The mean 
sky spectra obtained with the sky fibers were subtracted from the 
target spectra using {\sl skysub} in the {\sl imred} package.  This 
procedure was generally sufficient for the MMDOs.  The HPNe, however, 
were immersed in the inner bulge where a substantial surface 
brightness gradient exists.  Therefore, we conducted a ``second 
subtraction" procedure to deal with the inhomogeneous stellar 
background, which will be further addressed in Section~\ref{sec:line}.

\begin{figure*}\centering
\includegraphics[scale=1.,angle=0]{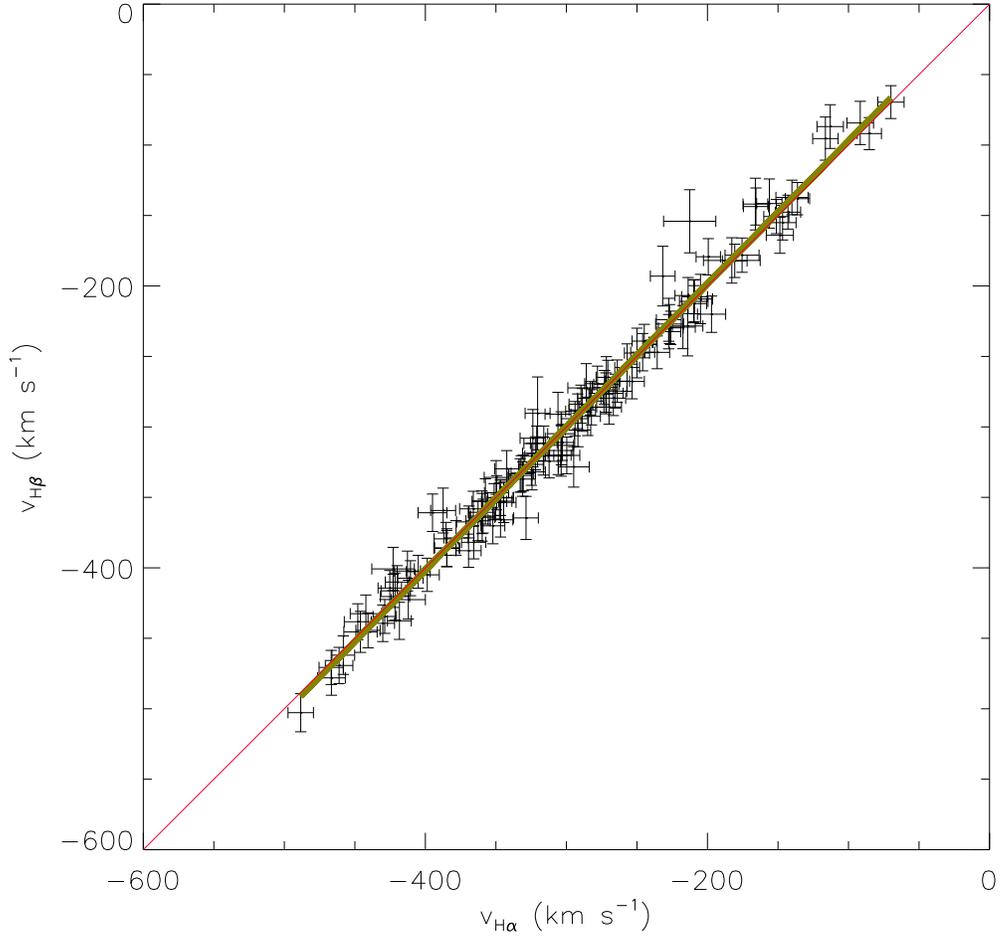}
\caption{A comparison between the H$\alpha$ and H$\beta$ radial 
velocities of MMDOs derived from our WIYN/Hydra dataset. The two 
lines are fitted independently with two Gaussian lines.  Only 
objects with S/N$>$3 in both lines are included. The thin red 
diagonal line represents a 1:1 correspondence.  The thick green 
line is the best linear-fit, which indicates no systematic offset 
between the two velocities and thus verifies the accuracy of our 
wavelength calibration.}
\label{fig:reffvhahb}
\end{figure*}

We tested the accuracy of our wavelength calibration by comparing 
the centroids of H$\alpha$ and H$\beta$ lines in the spectra of the 
MMDOs, each fitted independently with a Gaussian profile 
(Section~\ref{sec:line}).  As illustrated in 
Figure~\ref{fig:reffvhahb}, the derived radial velocities of 
H$\alpha$ and H$\beta$ are in excellent agreement, with an average 
offset of $0.04{\rm~km~s^{-1}}$ and an root-mean-square (rms) of 
$13.1{\rm~km~s^{-1}}$. Throughout this work, radial velocities have 
been corrected to heliocentric values. 

\section{Analysis and results} \label{sec:result}

We analyze the Hydra spectra of the 77 HPNe and 300 MMDOs, focusing 
on the most prominent nebular emission lines.  We first describe the
procedure to fit the lines in Section~\ref{sec:line}, and then 
present the results of emission line diagnostics 
(Section~\ref{sec:ratios}) and kinematics 
(Section~\ref{sec:kinematics}). 

\subsection{Emission line fitting} \label{sec:line}

\begin{figure*}\centering
\includegraphics[width=13cm,angle=0]{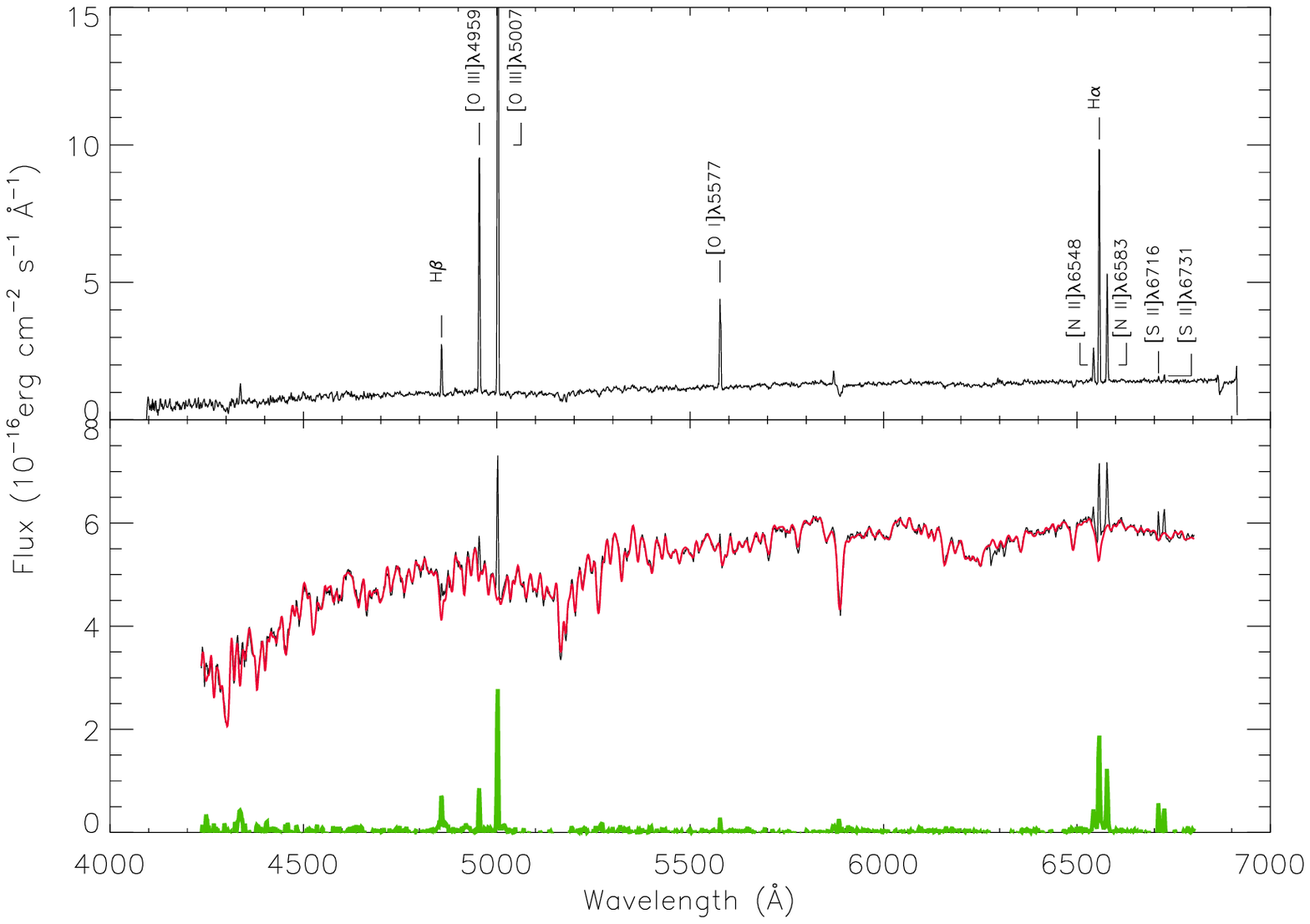}
\caption{{\it Top panel}: A typical MMDO spectrum (MMDO\,907), 
showing eight prominent emission lines: H$\beta$, [O~{\sc iii}] 
$\lambda\lambda$4959,5007, [N~{\sc ii}] $\lambda\lambda$6548,6583, 
H$\alpha$, and [S~{\sc ii}] $\lambda\lambda$6716,6731.  Note the 
residual [O~{\sc i}] $\lambda$5577 sky emission line, which is not 
included in our analysis.  {\it Bottom panel}: A typical Hydra 
spectrum of HPNe (\#63) is shown as the black curve.  The continuum 
is dominated by bulge stellar continuum emission falling in the 
fiber.  Red curve is the best-fit stellar continuum using the 
pPXF+MILES synthetic library.  The green curve shows the residual 
PN spectrum after stellar subtraction, where all eight emission 
lines remain.} 
\label{fig:reffdiskpnspec}
\end{figure*}

We begin with the MMDOs, the majority of which are located in the 
M31 disk and virtually unaffected by the strong stellar background 
emission.  A typical MMDO spectrum is demonstrated in the top panel 
of Figure~\ref{fig:reffdiskpnspec}. Eight emission lines are clearly
shown, including H$\beta$, [O~{\sc iii}] $\lambda\lambda$4959,5007, 
[N~{\sc ii}] $\lambda\lambda$6548,6583, H$\alpha$ and [S~{\sc ii}] 
$\lambda\lambda$6716,6731.  We fit the lines in the following steps.

First, we assume that each emission line has the Gaussian profile, 
and that any residual continuum after sky subtraction can be 
accounted for by a piecewise power-law in the vicinity of the lines.
The Gaussian centroid of the line defines the radial velocity, and 
the observed width of the Gaussian profile is determined by the 
convolution of the intrinsic dispersion ($\sigma_{\rm PN}$) and the 
instrumental broadening ($\sigma_{\rm inst} \approx$1.41\AA, or 
$\sim$84~km\,s$^{-1}$ at 5000\,\AA). 
Second, we tie the radial velocities and $\sigma_{\rm PN}$ among 
H$\beta$, H$\alpha$ and [N~{\sc ii}] $\lambda\lambda$6548,6583. 
Visual examination of the spectra indicates that this is reasonable 
for all cases.  The parameters of the [S~{\sc ii}] doublet, which 
have relatively low signal-to-noise ratio (S/N), are also tied to 
those of H$\alpha$.  [O~{\sc iii}] $\lambda\lambda$4959,5007 are of 
higher ionization and might arise from regions with different 
physical conditions in the nebula; they are fitted with an 
independent radial velocity and $\sigma_{\rm PN}$. Third, we tie the 
intensities of the [O~{\sc iii}] lines according to the theoretical 
ratio of $\lambda$4959/$\lambda$5007 =1:3.  The same ratio was 
assumed for the [N~{\sc ii}] doublet.  In total, we have 10 free 
parameters for eight emission lines, plus several parameters for the 
power-law continuum, which are of less interest and will not be 
further discussed here. 

\begin{figure*}\centering
\includegraphics[width=13cm,scale=1,angle=0]{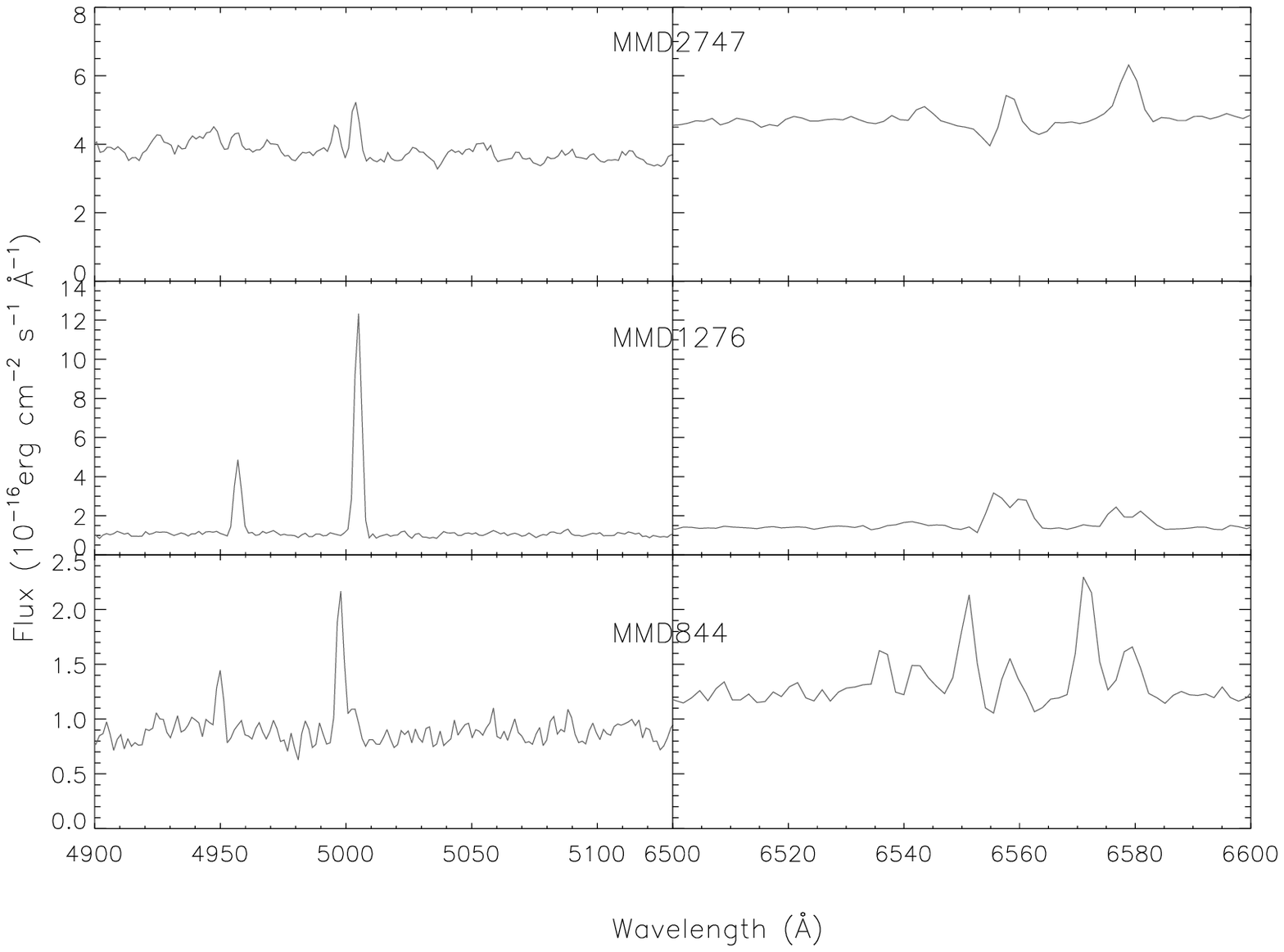}
\caption{Three MMDOs exhibiting double-component emission lines. 
MMDO\,844 has two components in [O~{\sc iii}], [N~{\sc ii}] and 
H$\alpha$.  MMDO\,1276 has double-peak in H$\alpha$ and [N~{\sc 
ii}].  MMDO\,2747 has a double-peak profile in [O~{\sc iii}]. 
}
\label{fig:reffdoublecom}
\end{figure*}

In three cases, a single Gaussian profile provides a poor fit to the 
lines, in particular [O~{\sc iii}] $\lambda$5007 and H$\alpha$. 
Visual examination indicates that this is probably due to the 
presence of a second velocity component, as illustrated in 
Figure~\ref{fig:reffdoublecom}.   Hence we add a second Gaussian 
component for such cases, linking its intrinsic dispersion to the 
primary component.  The spectra are fitted using the {\sc idl} 
routine {\sl mpfitexpr} \citep{Bevington92}.  The derived emission 
line intensities, radial velocities and intrinsic dispersions of 
the MMDOs are given in Table~\ref{diskpntable}.  Errors for these 
parameters are estimated using a bootstrapping method.  We corrected 
the line intensities for the Galactic foreground extinction, using 
an average Galactic extinction curve and $E$(B-V)=0.062 
\citep{Schlafly11}.

\begin{landscape}
\begin{deluxetable}{ccccccccccccccccccccc}
\tablecaption{Emission lines of MMDOs}
\tabletypesize{\tiny}
\tablewidth{0pt}
\tablehead{
\colhead{Object} &
\colhead{RA} &
\colhead{DEC} &
\colhead{v$_{H\alpha}$} &
\colhead{v$_{[O III]}$} &
\colhead{$\sigma_{H\alpha}$} &
\colhead{$\sigma$$_{[O III]}$} &
\colhead{F$_{H\beta}$} &
\colhead{F$_{[O III]}$} &
\colhead{F$_{H\alpha}$} &
\colhead{F$_{[N II]}$} &
\colhead{F$_{[S II]6716}$} &
\colhead{F$_{[S II]6731}$}\\
\colhead{(1)} &
\colhead{(2)} &
\colhead{(3)} &
\colhead{(4)} &
\colhead{(5)} &
\colhead{(6)} &
\colhead{(7)} &
\colhead{(8)} &
\colhead{(9)} &
\colhead{(10)} &
\colhead{(11)} & 
\colhead{(12)} &
\colhead{(13)} \\
}
\startdata

MMD1062&      10.30333&       41.22556&  -556.$\pm$         1.3&   -556.$\pm$         1.1&   25.$\pm$         1.5&     0.$\pm$         0.0& 0.17$\pm$0.01&2.10$\pm$0.05&0.79$\pm$0.03&0.07$\pm$0.01&0.00$\pm$0.00&0.01$\pm$0.01&\\ \\
MMD1053&      10.32167&       41.26694&  -441.$\pm$         1.3&   -453.$\pm$         1.1&    0.$\pm$         0.0&    13.$\pm$         0.9& 0.15$\pm$0.02&2.42$\pm$0.04&0.62$\pm$0.02&0.05$\pm$0.01&0.00$\pm$0.01&0.02$\pm$0.01&\\ \\
MMD1039&      10.32292&       41.20167&  -370.$\pm$         0.4&   -367.$\pm$         0.8&    0.$\pm$         0.0&     0.$\pm$         0.0& 0.13$\pm$0.01&2.45$\pm$0.04&0.92$\pm$0.01&1.58$\pm$0.02&0.12$\pm$0.01&0.12$\pm$0.01&\\ \\
MMD2732&      10.32333&       41.31250&  -404.$\pm$         7.8&   -389.$\pm$         3.8&   28.$\pm$         9.9&    27.$\pm$         6.7& 0.05$\pm$0.01&0.44$\pm$0.04&0.18$\pm$0.03&0.04$\pm$0.01&0.00$\pm$0.01&0.00$\pm$0.00&\\ \\
MMD1576&      10.33292&       41.13917&  -433.$\pm$         0.2&   -433.$\pm$         0.8&    0.$\pm$         0.0&     0.$\pm$         0.0& 3.33$\pm$0.05&3.53$\pm$0.05&22.99$\pm$0.10&5.81$\pm$0.03&1.67$\pm$0.04&1.15$\pm$0.05&\\ \\
MMD1029&      10.33625&       41.17139&  -296.$\pm$         0.8&   -288.$\pm$         1.0&    0.$\pm$         0.0&     0.$\pm$         0.0& 0.23$\pm$0.02&2.66$\pm$0.06&1.46$\pm$0.03&0.61$\pm$0.02&0.04$\pm$0.00&0.08$\pm$0.01&\\ \\
MMD784&      10.34000&       41.34500&  -324.$\pm$         0.2&   -323.$\pm$         0.8&    0.$\pm$         0.0&     0.$\pm$         0.0& 2.90$\pm$0.07&4.08$\pm$0.06&12.60$\pm$0.08&2.99$\pm$0.03&1.09$\pm$0.02&0.74$\pm$0.02&\\ \\
MMD1051&      10.34750&       41.24111&  -403.$\pm$         0.4&   -397.$\pm$         0.6&    0.$\pm$         0.0&     0.$\pm$         0.0& 0.65$\pm$0.02&12.50$\pm$0.11&4.91$\pm$0.05&1.55$\pm$0.02&0.04$\pm$0.01&0.12$\pm$0.01&\\ \\
MMD2787&      10.37958&       41.16000&  -433.$\pm$         0.7&   -427.$\pm$         0.7&    0.$\pm$         0.0&     0.$\pm$         0.0& 0.34$\pm$0.02&5.08$\pm$0.06&1.20$\pm$0.03&1.04$\pm$0.02&0.06$\pm$0.01&0.08$\pm$0.02&\\ \\
MMD1044&      10.39042&       41.21556&  -402.$\pm$         0.7&   -403.$\pm$         0.8&    0.$\pm$         0.0&     0.$\pm$         0.0& 0.27$\pm$0.02&5.43$\pm$0.09&1.68$\pm$0.04&0.73$\pm$0.02&0.06$\pm$0.02&0.00$\pm$0.00&\\ \\
\enddata

\tablecomments{(1) name of MMDOs, numbered as in \cite{Merrett06a}; (2)-(3) Sky coordinates, in Epoch 2000; (4) Heliocentric radial velocity of the H$\alpha$ emission line, in units of km~s$^{-1}$; 
(5) Heliocentric radial velocity of [O III]$\lambda$4959,5007, in units of km~s$^{-1}$; 
(6) Intrinsic velocity dispersion of H$\alpha$, in units of km s$^{-1}$; (7) Intrinsic velocity dispersion of [O III]$\lambda$4959,5007, in units of km~s$^{-1}$. 
Values of zero in column (6)-(7) indicate objects whose emission lines appear narrower than the instrument broadening. 
(8)-(13) Fluxes of the emission lines, in units of $10^{-15}{\rm~erg~s^{-1}~cm^{-2}}$.
For three objects exhibiting double velocity components, the velocities and fluxes of both components are reported.  
Only a portion of the full table is shown here to illustrate its form and content. The full table will be published online. 
}
\label{diskpntable}
\end{deluxetable}
\end{landscape}

The situation for the HPNe is more complicated, due to contamination 
from the stellar bulge and the existence of diffuse ionized gas in 
the bulge \citep{Jacoby85,Ciardullo88}.  The spectra of our target 
PNe are dominated by the continuum emission of bulge stars, as shown
in an example spectrum in the bottom panel of 
Figure~\ref{fig:reffdiskpnspec}.  Therefore, we employ the Penalized 
Pixel-Fitting (pPXF; \citealp{Cappellari04,Cappellari17}) package to 
fit and subtract the stellar continuum.  For the synthetic stellar 
template, we adopt the MILES library \citep{Falcon11}. 
pPXF decomposes the fitted spectrum into simple stellar populations 
(SSPs) with different ages and metallicities.  A byproduct of the 
fit is the integrated stellar velocity field, in particular the 
line-of-sight velocity ($V_*$) and velocity dispersion 
(${\sigma}_*$).  By design, pPXF masks out the narrow wavelength 
ranges around the potential emission lines, e.g., H$\alpha$.  The 
Balmer emission lines, especially H$\beta$, of the PNe would 
be affected by the Balmer absorption lines present in most SSPs. 
The forbidden lines, on the other hand, are free of this problem. 
We proceed with this caveat in mind and further address the 
potential effect in Section~\ref{sec:ratios}. 

Since the current wavelength range (4119--6882\,{\AA}) does not 
cover the Balmer break at 3646\,{\AA} \citep{Kauffmann03} or the 
Ca~{\sc ii} K and H lines \citep{Wilson70}, our spectra are 
insensitive to the intermediate-age ($\lesssim$ 1~Gyr) SSPs, which 
generally exhibit strong Balmer absorption lines. 
Fortunately, such intermediate-age SSPs, if any, are unlikely to be 
substantial in the M31 bulge (\citealt{Saglia10};\citealt{Dong15}). 
Therefore, we select a subset of the MILES library to fit the 
spectra, which include SSPs with ages from 1 to 14.125~Gyr and 
metallicity $Z$=0.008--0.033.  This choice is warranted by our 
analysis of the stellar populations based on a CAHA/PPAK 
integral-field spectroscopic mapping of the M31 bulge 
(R.~Garc\'{i}a-Benito et al.\ 2018, in preparation), which cover the 
Ca~{\sc ii} K and H lines as well as the Balmer break.  The hence 
fitted SSPs are predominantly old and are of 1--1.5 solar 
metallicity, which is consistent with the finding of \citet{Saglia10}
based on the long-slit spectra.  It is noteworthy that the enclosed 
stellar mass within each 2$\arcsec$-diameter fiber is estimated to 
be 10$^{5}$--10$^{6}$\,$M_{\odot}$, based on the projected stellar 
mass distribution derived in \cite{Dong15}, and thus stochastic 
fluctuation due to a few bright stars in the modeled SSPs is 
negligible. 
An example of the modeled stellar continuum is shown by a red curve 
in the bottom panel of Figure~\ref{fig:reffdiskpnspec}.

After subtraction of the modeled stellar continuum, we employ the same approach as for the MMDOs to fit the emission lines, which should predominantly arise from the HPNe, although the presence of diffuse ionized gas might cause some contamination, due to the limitation of our multi-fiber observations (Section ~\ref{sec:detection}). We assess the degree of contamination from the diffuse ionized gas using our \emph{HST} WFC3 narrow-band images (Figure ~\ref{fig:bulgemap}; Z. Li et al. in preparation; see also \citealt{Dong14,Dong16}).  It can be seen from Figure ~\ref{fig:bulgemap} that the majority of the HPNe are located outside the so-called {\it nuclear spiral} \citep{Jacoby85}, where the diffuse ionized gas is concentrated.  Using the [O III] image and taking into account the sky area covered by the fibers, we find that, in 52 out of the 77 HPNe, the diffuse ionised gas contributes less than 20\% (and typically less than 10\%) of the measured [O III] flux.  This number drops to 51 HPNe, if we required that the contamination is less than 20\% in all three lines of [O III], H$\alpha$, and [N II].  We restrict the following statistical analysis to this subsample of 51 HPNe.

It turns out that all 77 HPNe exhibit the [O~{\sc iii}] $\lambda$5007
line with S/N$>$5, confirming that they are genuine [O~{\sc 
iii}]-emitting objects. We also find that a second velocity component
is required for three objects (\#21, \#45 and \#52), which exhibit 
double peaks in [O~{\sc iii}], and a fourth object (\#50) with 
double components in both [N~{\sc ii}] and H$\alpha$. The fit results
for the 77 HPNe, including line intensity, radial velocity and 
intrinsic dispersion, are given in Table~\ref{bulgepntable}. 

\begin{landscape}
\begin{deluxetable}{ccccccccccccccccccccc}
\tablecaption{Emission lines of HPNe}
\tabletypesize{\tiny}
\tablewidth{0pt}
\tablehead{
\colhead{Object} &
\colhead{RA} &
\colhead{DEC} &
\colhead{v$_{H\alpha}$} &
\colhead{v$_{[O III]}$} &
\colhead{$\sigma_{H\alpha}$} &
\colhead{$\sigma$$_{[O III]}$} &
\colhead{F$_{H\beta}$} &
\colhead{F$_{[O III]}$} &
\colhead{F$_{H\alpha}$} &
\colhead{F$_{[N II]}$} &
\colhead{F$_{[S II]6716}$} &
\colhead{F$_{[S II]6731}$}\\
\colhead{(1)} &
\colhead{(2)} &
\colhead{(3)} &
\colhead{(4)} &
\colhead{(5)} &
\colhead{(6)} &
\colhead{(7)} &
\colhead{(8)} &
\colhead{(9)} &
\colhead{(10)} &
\colhead{(11)} & 
\colhead{(12)} &
\colhead{(13)} \\
}
\startdata
HPN1&      10.64692&       41.26445&  -546.$\pm$         5.8&   -519.$\pm$        15.0&   96.$\pm$         8.4&    75.$\pm$        16.0& 0.40$\pm$0.04&0.35$\pm$0.05&0.63$\pm$0.05&0.37$\pm$0.07&0.07$\pm$0.02&0.08$\pm$0.02&\\ \\
HPN2&      10.65542&       41.27444&  -333.$\pm$         3.8&   -323.$\pm$        11.0&   55.$\pm$         5.7&    82.$\pm$         8.0& 0.56$\pm$0.07&1.24$\pm$0.11&1.58$\pm$0.11&1.28$\pm$0.09&0.80$\pm$0.07&0.58$\pm$0.07&\\ \\
HPN3&      10.65592&       41.26389&  -177.$\pm$         3.6&   -178.$\pm$         1.2&   75.$\pm$         7.3&    38.$\pm$         9.8& 1.99$\pm$0.12&14.47$\pm$0.83&3.04$\pm$0.22&0.21$\pm$0.07&0.02$\pm$0.06&0.06$\pm$0.07&\\ \\
HPN4&      10.65750&       41.26333&  -366.$\pm$         8.8&   -406.$\pm$        19.5&   56.$\pm$        27.2&    75.$\pm$        25.8& 0.31$\pm$0.07&0.52$\pm$0.10&0.87$\pm$0.21&0.54$\pm$0.14&0.09$\pm$0.05&0.15$\pm$0.05&\\ \\
HPN5&      10.65846&       41.26694&  -381.$\pm$        10.2&   -177.$\pm$        31.8&   62.$\pm$        19.4&   175.$\pm$        25.6& 0.69$\pm$0.11&1.64$\pm$0.29&0.83$\pm$0.17&1.30$\pm$0.25&0.70$\pm$0.11&0.60$\pm$0.11&\\ \\
HPN6&      10.65963&       41.25917&  -374.$\pm$         4.2&   -383.$\pm$         1.3&   62.$\pm$        10.1&    57.$\pm$         7.0& 0.98$\pm$0.10&6.91$\pm$0.35&2.91$\pm$0.27&0.54$\pm$0.11&0.21$\pm$0.08&0.18$\pm$0.07&\\ \\
HPN7&      10.66258&       41.26305&  -398.$\pm$        18.9&   -320.$\pm$        29.3&   64.$\pm$        26.1&   142.$\pm$        38.1& 0.23$\pm$0.06&0.57$\pm$0.14&0.31$\pm$0.10&0.35$\pm$0.10&0.14$\pm$0.06&0.10$\pm$0.06&\\ \\
HPN8&      10.66271&       41.27667&  -286.$\pm$         8.1&   -266.$\pm$         6.3&   91.$\pm$        16.6&    68.$\pm$        34.0& 0.73$\pm$0.09&1.58$\pm$0.38&1.44$\pm$0.20&1.13$\pm$0.16&0.45$\pm$0.08&0.28$\pm$0.06&\\ \\
HPN9&      10.66308&       41.27611&  -377.$\pm$         9.1&   -382.$\pm$        10.3&   74.$\pm$        18.1&    89.$\pm$        16.3& 0.48$\pm$0.07&0.96$\pm$0.12&0.67$\pm$0.12&0.62$\pm$0.12&0.21$\pm$0.07&0.18$\pm$0.06&\\ \\
\enddata

\tablecomments{(1) Name of HPNe; (2)-(13) The same as in Table~\ref{diskpntable}.
Only a portion of the full table is shown here to illustrate its form and content. The full table will be published online. 
}
\label{bulgepntable}
\end{deluxetable}
\end{landscape}

\subsection{Line intensity diagnostics} \label{sec:ratios} 
\begin{figure}\centering
\includegraphics[width=13cm,scale=1,angle=0]{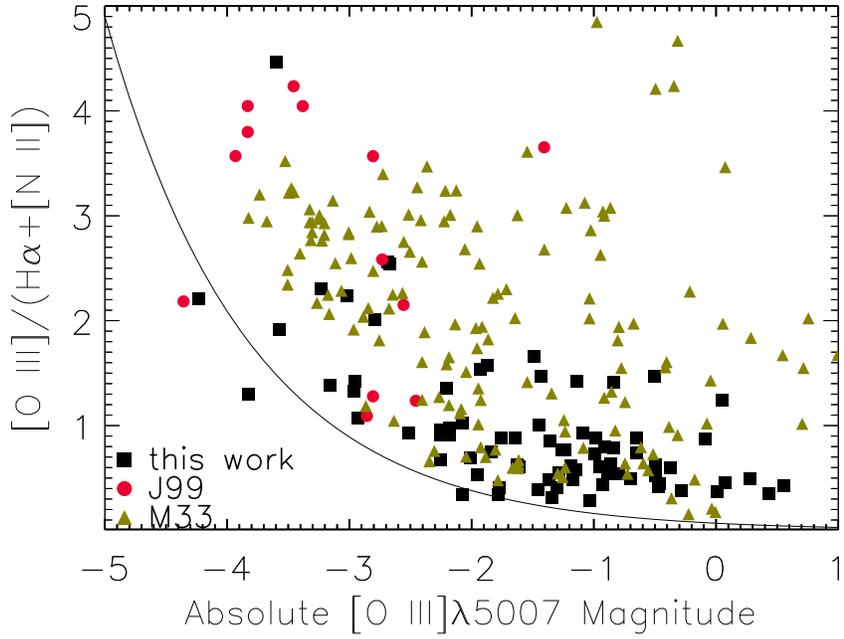}
\caption{The [O~{\sc iii}]$\lambda$5007/(H$\alpha$+[N~{\sc ii}])
ratio versus the [O~{\sc iii}] absolute magnitude.  Black squares: 
HPNe in this work; Red circles: PNe in M31's bulge \citep{Jacoby99};
Green triangles: PNe in M33 \citep{Ciardullo04}.  Black curve is 
the relation between the two quantities, in the form of 
$\log$X = $-$0.37$\times$Y $-$ 1.14 (see text for explanation).} 
\label{fig:reffcone}
\end{figure}

In this section we focus on the four brightest lines: H$\beta$, 
[O~{\sc iii}] $\lambda$5007, H$\alpha$, and [N~{\sc ii}] 
$\lambda$6583. Figure~\ref{fig:reffcone} displays the line flux ratio of $\rm R=\frac{[O\,III]}{H\alpha+[N\,II]}$ vs. $\rm M_{\lambda5007}$ for the 77 PNe. \cite{Herrmann08} and \cite{Ciardullo10} have shown that this diagram can be used to exclude contaminations in PNe 
identification. Specifically, genuine PNe satisfy the relation 
4$\textgreater$ log$R$ $\textgreater-0.37 \times M_{\lambda5007} - 
1.16$. As shown 
in Figure~\ref{fig:reffcone}, most of the HPNe meet this criteria, 
and it is also noteworthy that the overall distribution of HPNe in 
this diagram resembles that of the PNe from the literature, which also suggests that our targets are genuine PNe. 

\begin{figure}\centering
\includegraphics[scale=1,angle=0]{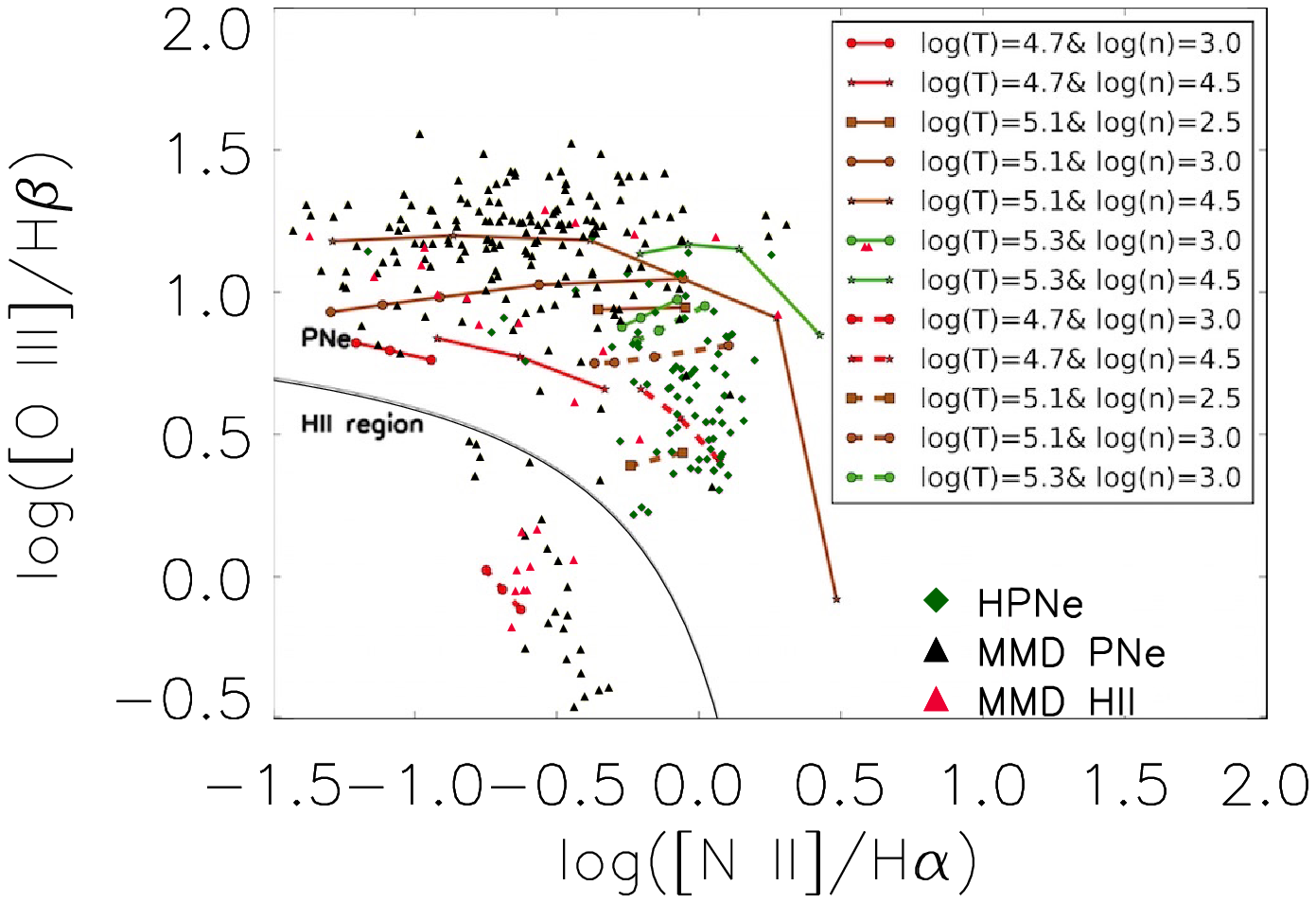}
\caption{A BPT diagram for HPNe and MMDOs. Green diamonds: HPNe candides; Black triangles: MMD PNe candidates; Red triangles: MMD HII region candidates. 
For the few objects with double-peak emission lines, the line flux is the sum of the two components. 
The black curve shows an empirical division of PNe and HII regions \citep{Sanders12}.
Squares circles, stars, and triangles linked by solid or dashed lines represent the 3MdB photoionization models (\citealp{Morisset15}) of the same effective temperature (T, in units of K) and column density (n, in units of cm$^{-3}$), but of different luminosities of the central star. 
Solid lines are for models of near-solar abundance (12+log[O/H]=8.64), mass-bounded model, while dashed lines are for ``very high" abundance (12+log[O/H]=9.24), radiation-bounded model.} 
\label{fig:reffbpt}
\end{figure}

We also construct a BPT diagram \citep{Baldwin81} for the 77 HPNe and 
300 MMDOs, as shown in Figure~\ref{fig:reffbpt}.  Following 
\cite{Sanders12}, we adopt an empirical curve to distinguish the 
H~{\sc ii} regions and PNe in this diagram.  All of HPNe (green 
diamonds) lie in the expected PN 
region.  The majority of the MMD PNe are also located in this 
region. However, 23 outliers might be actually H~{\sc ii} regions,
which needs further assessment.

No extinction was made when we measured these line ratios. However,
this is unlikely to affect the above line ratios, given the close 
wavelengths of the two emission lines in each ratio. On the other hand, as stated in Section~3.1, the measured values of 
H$\beta$ in the HPNe might be biased too high, due to degeneracy 
with the H$\beta$ absorption line as predicted by the modeled 
stellar spectra.  To correct for this potential bias in H$\beta$, 
we have set an upper-limit to the flux of H$\beta$ based on 
H$\alpha$.  Bright bulge PNe (which is the case for HPNe) generally 
have circumstellar extinction of around $A_{\lambda5007}\sim$0.8 
\citep{Richer99,Jacoby99}, and the inner bulge of M31 has been found
to have low local extinction \citep{Dong16}.  Thus by assuming the 
Case~B theoretical ratio H$\alpha$/H$\beta$=2.85 \citep{Storey95}, we estimate that $I$(H$\alpha$)/3.45 = $I$(H$\beta$).  This 
can be used as an upper-limit flux of H$\beta$ for the HPNe with 
initial H$\alpha$/H$\beta<$3.45.  We verify that the H$\beta$ 
intensity after extinction correction does not show any artificial 
radial trend.

The [N~{\sc ii}]$\lambda$6583/H$\alpha$ and 
[O~{\sc iii}]$\lambda$5007/H$\beta$ intensity ratios as a function of 
the projected galactocentric radius $R$ are shown in 
Figure~\ref{fig:intensity}, for the 51 HPNe in which contamination from diffuse emission is negligible (Section 3.1).  In each panel, only those objects (HPNe 
or MMDOs) with an S/N$>$3 in the four lines are included.  To see 
the trend of these ratios versus $R$, we also plotted the line 
ratios of M31 outer-disk and halo PNe retrieved from the literature 
\citep{Kwitter12,Balick13,Corradi15,Fang13,Fang15b,Fang18}.  The 
[O~{\sc iii}]/H$\beta$ ratios of our sample show larger dispersion,
especially for the HPNe. The HPNe, located exclusively
at $R<$0.5~kpc, show on average higher [N~{\sc ii}]/H$\alpha$ and 
lower [O~{\sc iii}]/H$\beta$ (Figure~\ref{fig:intensity}), compared 
to the MMDOs, which mostly reside at 0.5$<R<$4~kpc. The PNe in the outer-disk and halo 
of M31 generally have lower [N~{\sc ii}]$\lambda$6583/H$\alpha$ 
ratios than our sample of the HPNe but comparable [O~{\sc 
iii}]$\lambda$5007/H$\beta$ ratios. 

\begin{figure*}\centering
\label{fig:intensity}
\subfigure{
\label{fig:reffn2ha}
\includegraphics[width=2.7in,angle=0]{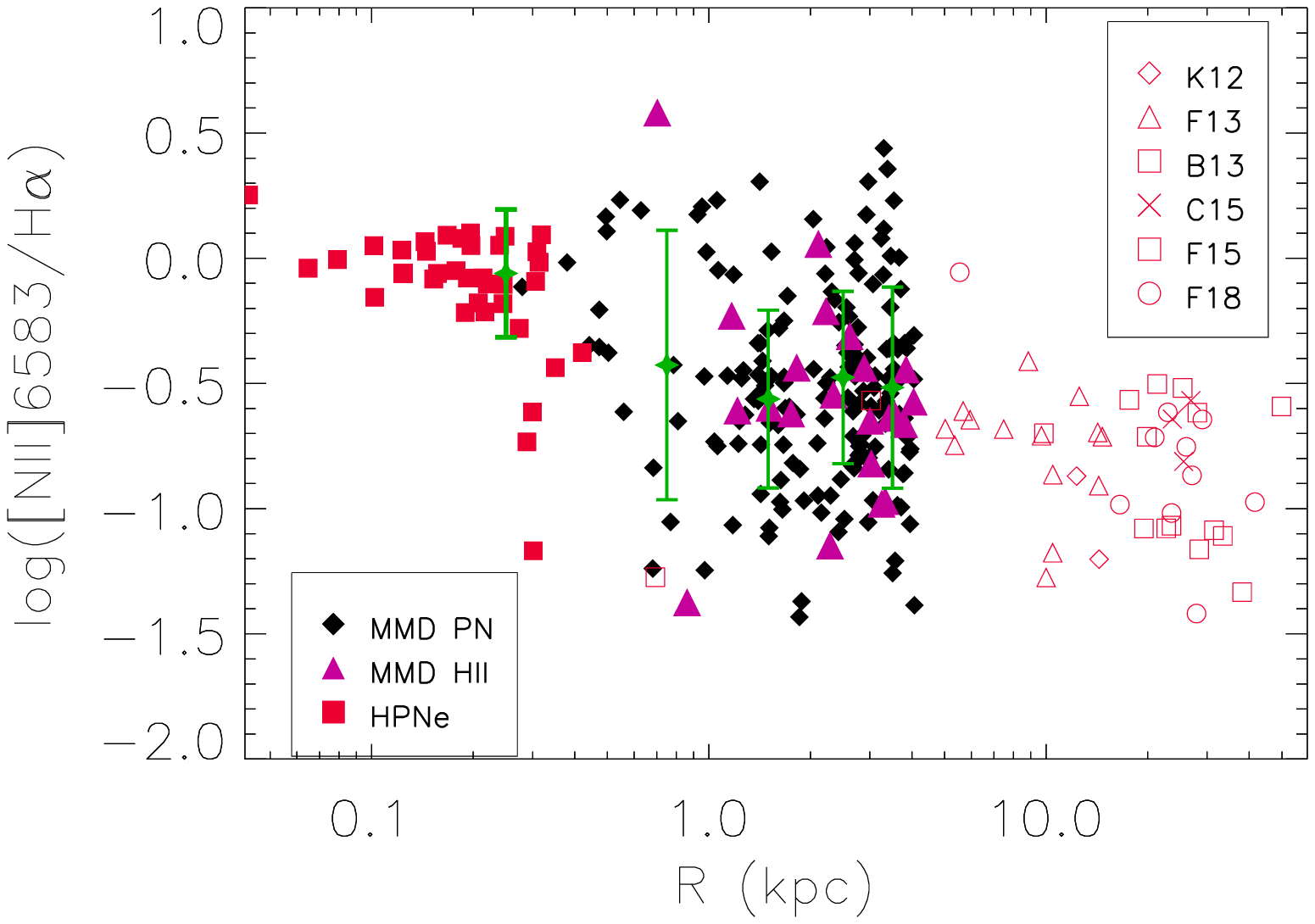}
\label{fig:reffn2ha}
}
\subfigure{
\label{fig:reffo3hb}
\includegraphics[width=2.7in,angle=0]{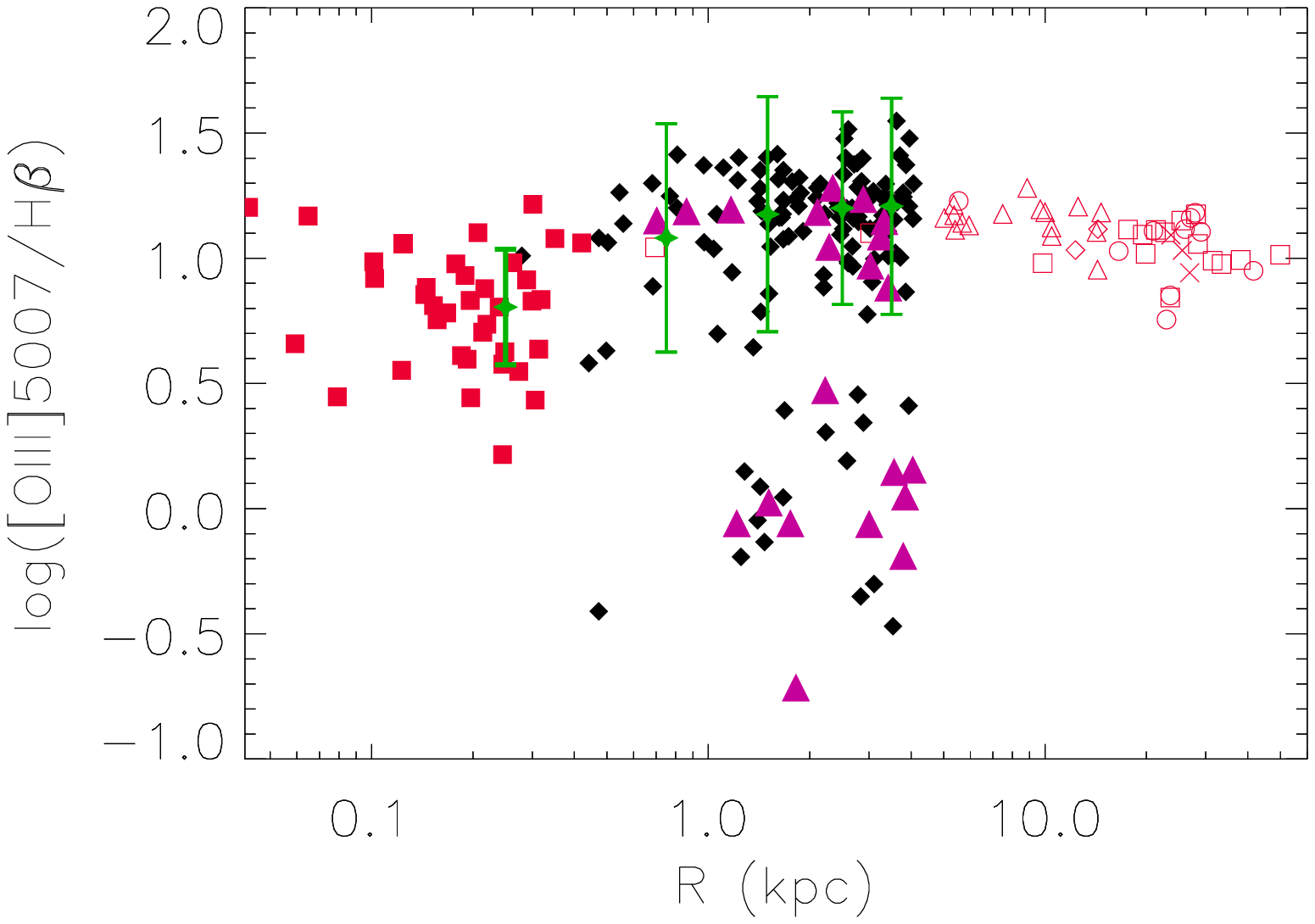}
\quad
}
\caption{[N~{\sc ii}]$\lambda$6583/H$\alpha$ (left) and [O~{\sc 
iii}]$\lambda$5007/H$\beta$ (right) intensity ratios as a function 
of projected galactocentric radius.  In both panels, black diamonds 
represent the MMD PNe, purple triangles are the MMD H~{\sc ii} 
regions, and red squares are the 51 HPNe in which contamination from diffuse emission is negligible (Section 3.1).  Only objects with S/N$>$3 
in these four emission lines are included. HPNe and the MMD PNe
are grouped by projected radii in 0--0.5, 0.5--1, 1--2, 2--3 and 
3--4 kpc; green dots are the median values of these five regions 
along with the 68\% quantile bar.  Red open symbols are the M31 
outer-disk and halo PNe retrieved from the literature (see text for 
description).}
\label{fig:intensity}
\end{figure*}

The higher [NII]/H$\alpha$ of HPNe can be naturally understood as the result of an on-average higher stellar nitrogen abundance in the inner bulge, compared to the disk (\citealt{Saglia10}; \citealt{Sanders12}).
The decreasing trend in [O III]/H$\beta$ can also be understood as a higher oxygen abundance, which in a photoionized gas would lead to a lower electron temperature and consequently a lower value of [O III]/H$\beta$\citep{Nagao06}. To further understand the occupation of HPNe in the BPT diagram, we explore the Mexican Million Models database (3MdB; \citealp{Morisset15}), which is a large set of numerical simulations for photoionized clouds (including PNe), based on CLOUDY \citep{Ferland98}. 
We plot in Figure~\ref{fig:reffbpt} model-predicted line ratios from the most relevant 3MdB PN models, with a variety of gas density and effective temperature of the central star, as well as metallicity.  
Apparently, the observed line ratios of both the HPNe and MMD PNe can be reproduced by the chosen models. In particular, the HPNe can be matched by models with a super-solar metallicity, while the MMD PNe can be matched by models with a near-solar metallicity.
This further supports a connection between the observed line ratio gradients and the metal-enriched bulge of M\,31.

\subsection{Kinematics} \label{sec:kinematics}

\begin{figure*}\centering
\subfigure[{           }]{
\includegraphics[width=2.75in,height=2.7in,angle=0]{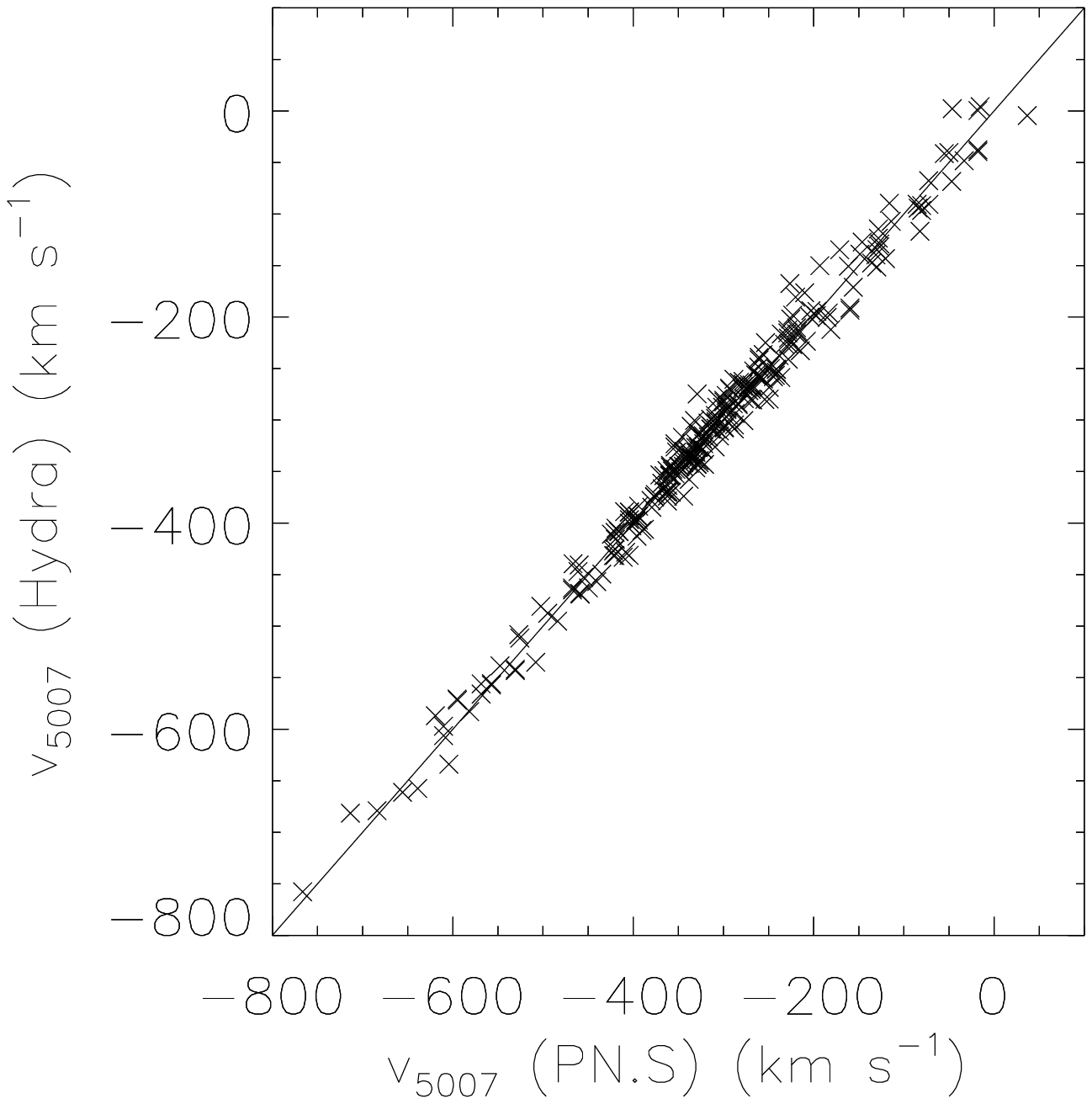}
\label{fig:reffvo3}
}
\subfigure[{      }]{
\includegraphics[width=2.75in,angle=0]{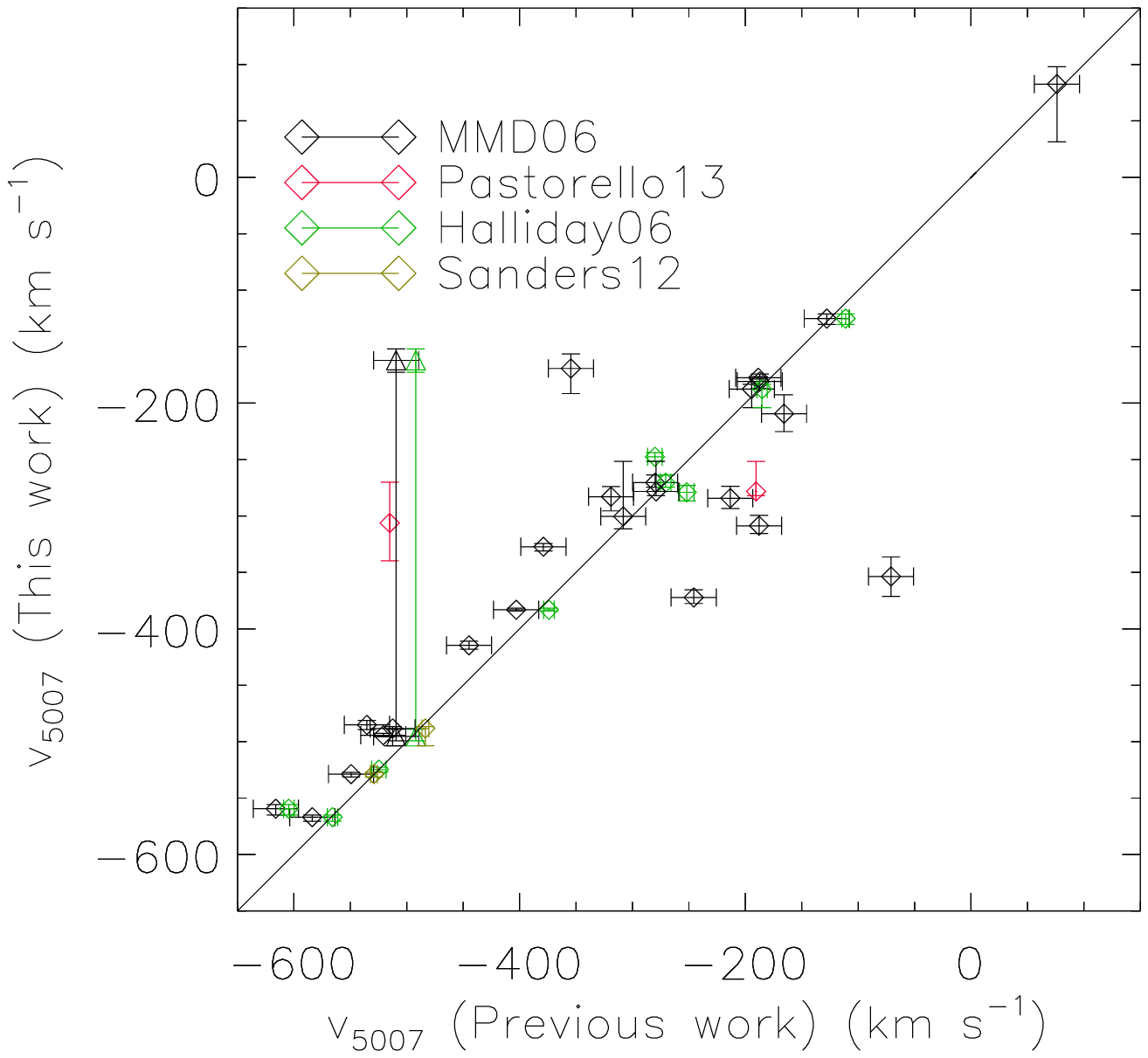}
\label{fig:reffcommer}
}
\caption{Comparison of [O~{\sc iii}] $\lambda$5007 radial velocities
measured from this work and from previous work. Left: MMDOs; Right: 
HPNe.  In the right panel, the black or green lines linking two 
triangles stand for the HPNe with double components in [O~{\sc iii}]
found in this work.  In both panels, the black diagonal line 
represents a 1:1 correspondence.} 
\label{fig:vcom}
\end{figure*}

For each object, we measured two radial velocities, one related to 
H$\alpha$ and the other related to [O~{\sc iii}] $\lambda$5007. 
Our measurements of the [O~{\sc iii}] velocities (only for the 
MMDOs) are compared with those derived by MMD06 in 
Figure~\ref{fig:vcom}-(a), and excellent agreement is seen. 
As for the 77 HPNe, 29 have at least one [O~{\sc iii}] velocity 
measurement in previous studies, of which 26 were reported by 
MMD06, 10 by \cite{Halliday06}, 2 by \cite{Sanders12}, and another 
2 by \cite{Pastorello13}. 
A comparison with these measurements is shown in 
Figure~\ref{fig:vcom}-(b).  While reasonable agreement is found 
for most objects, in several cases obvious difference still exists.

We suggest that our results should be more robust, due to our 
deeper exposures and extra use of the [O~{\sc iii}] $\lambda$4959 
nebular line.  It is noteworthy that the PNe with a second velocity 
component in [O~{\sc iii}] or H$\alpha$ (Section~\ref{sec:line}) 
generally have one component consistent with previous measurements.
The two components in the velocity have separations of $\gtrsim$
200 km\,s$^{-1}$.  This might signify fast bipolar outflows in 
PNe, which are highly inclined with respect to the line of sight. 
Bipolar outflows with expansion velocities as fast as 300 
km\,s$^{-1}$ have been detected in the Milky Way PNe (e.g., 
\citealt{Fang15,Danehkar16}).  The origin of such fast outflows 
remains elusive, but this topic is beyond the scope of discussion 
of this work.

\begin{figure*}\centering
\includegraphics[width=14cm,angle=0]{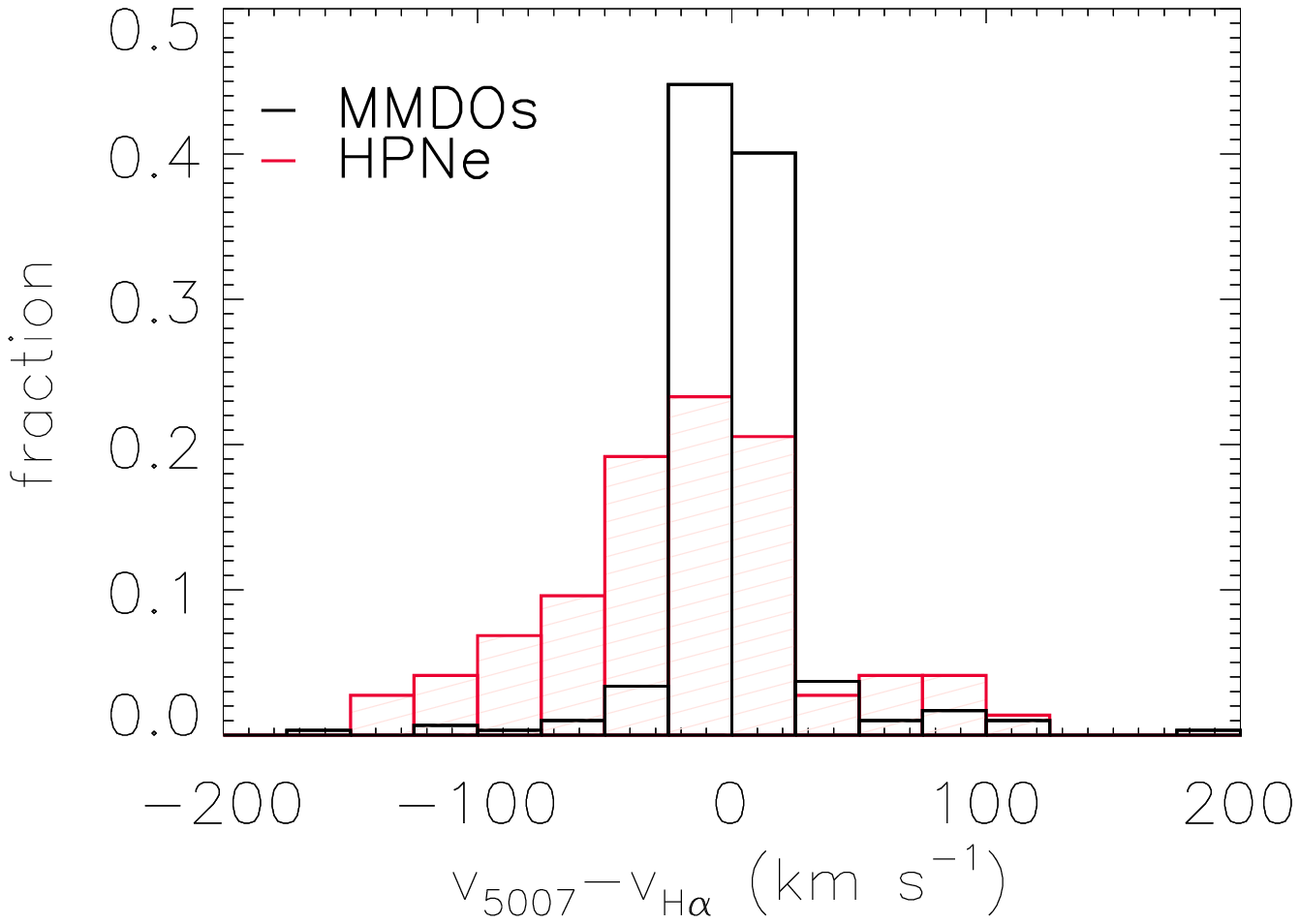}
\caption{Histograms of the difference between the radial velocities 
derived from [O~{\sc iii}] and H$\alpha$.}
\label{fig:reffo3-ha-hist}
\end{figure*}

We also check the distribution of the difference between the 
[O~{\sc iii}] and H$\alpha$ velocities, $V_{\lambda5007}$ $-$ 
$V_{{\rm H}\alpha}$, of the two groups (HPNe and MMDOs), as shown 
in Figure~\ref{fig:reffo3-ha-hist}.  For the MMDOs, there is no 
obvious systematic offset between $V_{\lambda5007}$ and 
$V_{{\rm H}\alpha}$, although the maximum difference approaches 
$\pm$200~km\,s$^{-1}$.  The HPNe show a somewhat larger scatter in 
this difference. 
For the HPNe, there is an extended distribution of the velocity 
difference in the range $-$160 to $-$100~km\,s$^{-1}$, but this 
asymmetry is statistically insignificant (only $\sim$2\,$\sigma$). 
[O~{\sc iii}] requires higher excitation, so its location in PNe 
and expansion velocity may be different from other emission lines. 
Unless otherwise specified, in the discussion hereafter, we use the 
line-of-sight velocity measured from H$\alpha$, whose emission 
distribution in PNe is more homogeneous than [O~{\sc iii}] and thus 
is expected to be more representative, to carry out the kinematic 
and dynamical analysis. 

\begin{figure*}\centering
\subfigure{
\includegraphics[scale=0.6,angle=0]{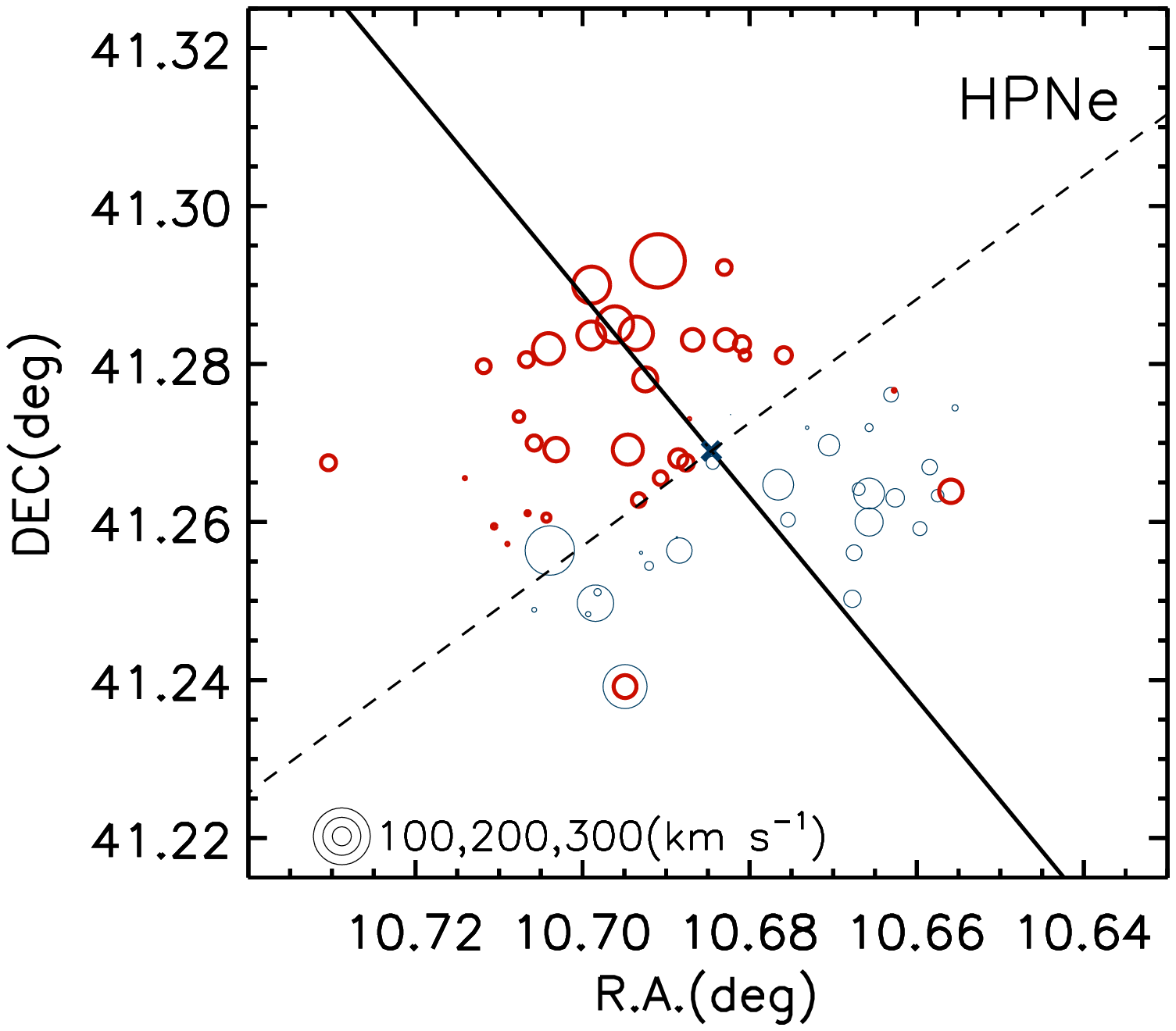}
\label{fig:reffbulgev}
}
\subfigure{
\label{fig:reffdiskv}
\includegraphics[scale=0.6,angle=0]{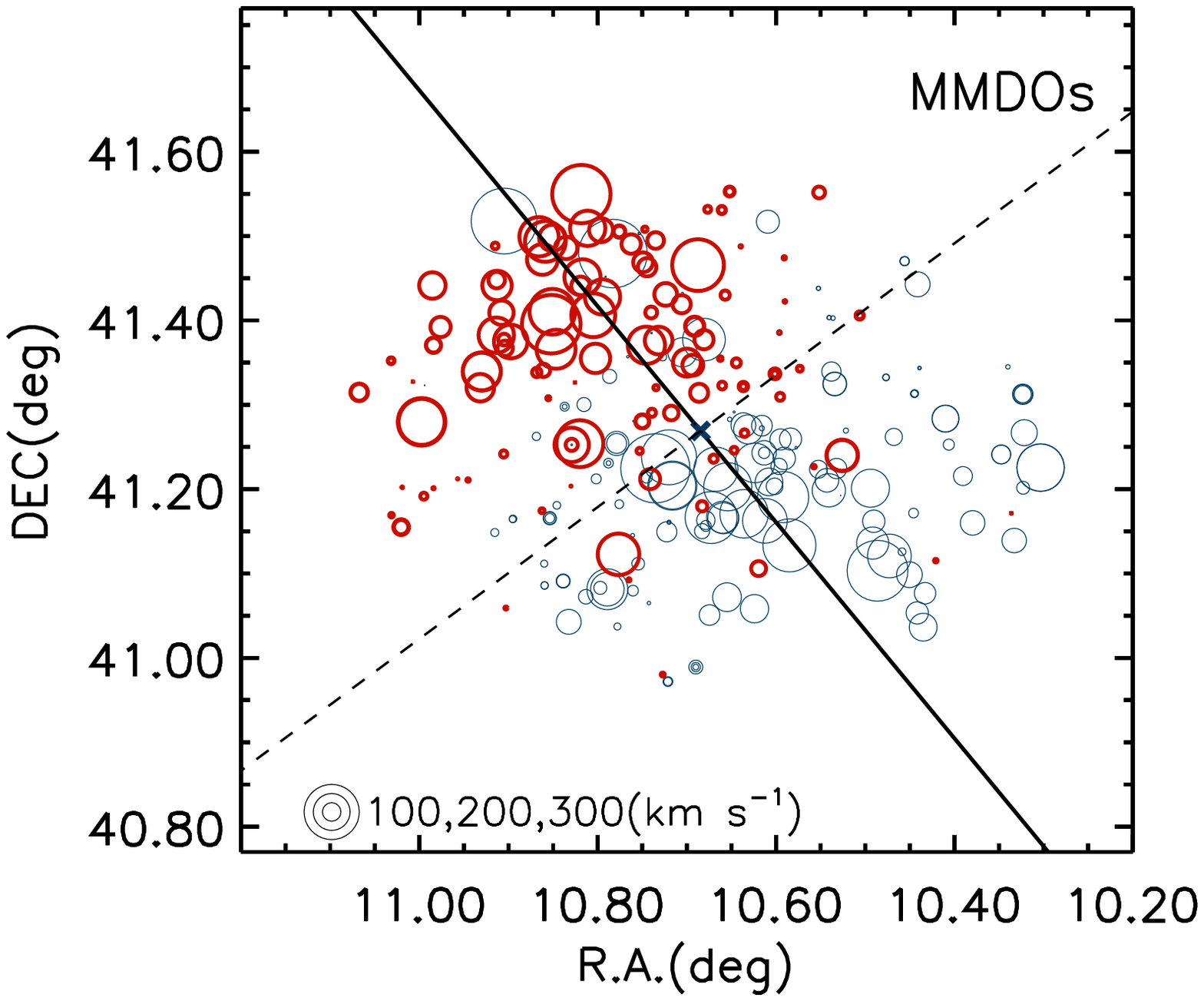}
\quad
}
\caption{Radial velocity (corrected for the systemic velocity of 
M31) map of the spectroscopically confirmed HPNe (top) and MMO PNe 
(bottom).  Red and blue circles represent red- and blue-shifted 
objects, respectively, with the circle size proportional to the 
absolute velocity.  The solid and dashed lines indicate the major- 
and minor-axis of M31, respectively.  A rotation pattern about the minor-axis is evident in both panels.} 
\label{fig:rotation}
\end{figure*}

After correcting for M31's systemic (heliocentric) velocity of 
$-$301 km\,s$^{-1}$ \citep{Vaucouleurs91}, the radial velocities 
$V_{{\rm H}\alpha}$ of HPNe range from $-$245~km\,s$^{-1}$ to 
283~km\,s$^{-1}$; 90\% of them lie within $\pm$166~km\,s$^{-1}$. 
This is compatible with the bulge velocity dispersion of 
$\sim$166~km\,s$^{-1}$ \citep{Saglia10}, which is a corroborative 
support to our presumption that all our HPNe reside in M31's bulge.
On the other hand, the MMDO objects have radial velocities between 
$-$374~km\,s$^{-1}$ and 319~km\,s$^{-1}$, which is compatible with 
the maximum circular velocity ($\sim$250~km\,s$^{-1}$; 
\citealt{Braun91}) of the M31 disk. 


Given that the velocity ranges of these two groups of PNe are 
consistent with the literature, we expect that their radial 
velocity distributions are also consistent with the intrinsic 
rotation pattern of M31 (here we assume a clockwise rotation of M31 
as observed from the top of the disk).  The spatial distribution 
of radial velocities in M31 disk is visually illustrated in 
Figure~\ref{fig:rotation}, where HPNe and MMDOs with the red-shifted
and blue-shifted velocities (with respect to the systemic velocity 
of M31) are displayed separately.  The radial velocity map is 
supposed to be symmetric about the minor axes of the disk where the 
signs of the radial velocities of the PNe distributed in the 
north and south regions are opposite. 

Indeed, the MMDOs show the radial velocities approximately symmetric
about the minor-axis of the disk (with a position angle of 
$\sim$128$^{\circ}$, east of south; \citealt{Kent89}): 
79$\pm$10\% of the MMDOs in the northeastern 
disk region are red-shifted, while 88$\pm$11\%  in 
the southwestern disk region are blue-shifted.  Interestingly, 
the HPNe display a similar rotation pattern: 97$_{-25}^{+3}$\%
of the HPNe  in the northeast  region are red-shifted; 
90$_{-24}^{+10}$\% of the HPNe  in the southwest region are blue-shifted. 
We further test this pattern by considering the position angle of 
the bulge kinematic minor axis (${\rm PA}_{\rm k}$) as a free 
parameter, which is derived by maximizing the sum of 
${\sum_{i}[V_i\,{\rm cos}({\rm PA}_i - {\rm PA}_{\rm k})]}$, where 
$V_i$ and ${\rm PA}_i$ are the radial velocity and position angle 
of the $i$th HPNe, respectively.  This results in $\rm PA_{\rm k}
=130^\circ\pm4.1^\circ$. 
$\rm PA_{\rm k}$ can also be estimated using the line-of-sight 
velocity of the integrated stellar continuum at the position of 
each HPN, as derived from our pPXF fits (Section~3.1).  This leads 
to $\rm PA_{\rm K} =125^\circ\pm9.2^\circ$.  Both values of $\rm 
PA_{\rm K}$ are consistent, within the errors, with the position 
angle of the disk minor-axis. 

Therefore, both discrete tracers (i.e., the HPNe) and the integrated
starlight show significant rotation in the inner bulge of M31, as 
previously found by \cite{Saglia10} through long-slit spectroscopy.
Nevertheless, \cite{Saglia10} argued that the M31 bulge is a slow 
rotator, which has an intensity-weighted mean $V/\sigma \approx
0.2$.  We also calculate an intensity-weighted average of 
$V/\sigma$, using $V_{*}$ and $\sigma_{*}$ measured at the position
of each HPNe; here the intensity refers to the median of the 
underlying stellar continuum in 5000--5500\AA. A similar value of 
$V/\sigma \approx 0.20$ is found, which confirms that the inner 
bulge of M31 is a slow rotator. 
\begin{figure*}\centering
\includegraphics[scale=0.7,angle=0]{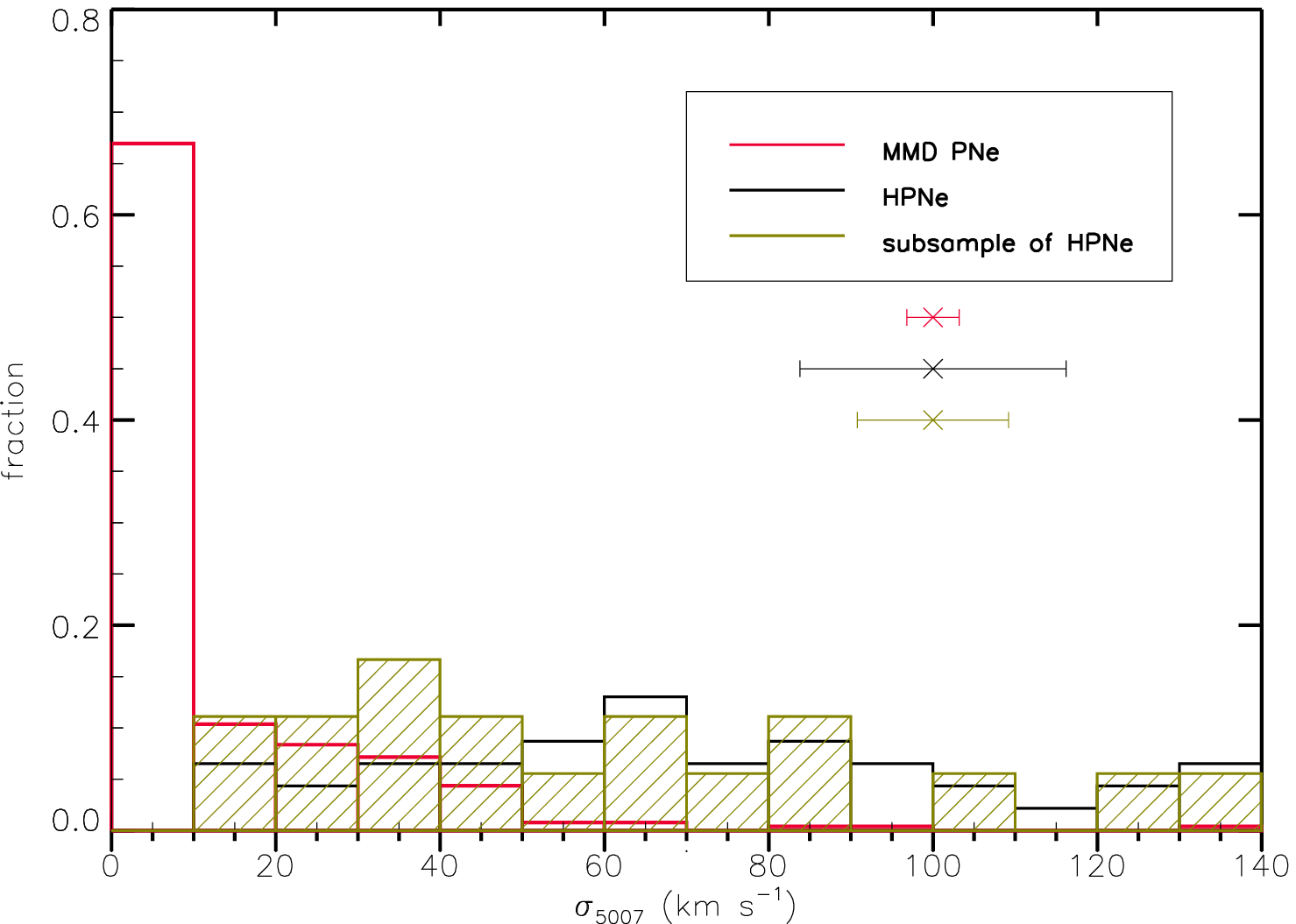}
\caption{The fractional distribution of the intrinsic dispersion of the [O III] line, for the MMD PNe (red), all HPNe (black), and a subset of HPNe with S/N greater than 10 in [O III] (green).}
\label{fig:sig_sl}
\end{figure*}

Lastly, we examine the distribution of the intrinsic dispersion of 
the [O III] line, $\sigma_{\rm 5007}$, as shown in 
Figure~\ref{fig:sig_sl}. 
Due to the moderate spectral resolution ($\sigma_{\rm inst} \approx 
84{\rm~km~s^{-1}}$) and typical PN expansion velocities of $
\lesssim$30${\rm~km~s^{-1}}$, the observed lines are not expected to 
be resolved. 
This is indeed the case for $\sim$82\% of the MMDOs, which have 
$\sigma_{\rm 5007}$ consistent with zero. On the other hand, 
$\sim$54\% of the HPNe exhibit $\sigma_{\rm 5007}$ greater than 
$70{\rm~km~s^{-1}}$. 
To see whether the bulge diffuse emission might have affected the 
measurement, we also plot in Figure~\ref{fig:sig_sl} a subset of 18 
HPNe (green histogram), whose [O III]$\lambda$5007 line is detected 
with an S/N $\geq$ 10, ensuring a robust measurement for the 
dispersion. This subset also shows a significant fraction of large 
$\sigma_{\rm 5007}$, suggesting that this is an intrinsic property.
We address the possible physical cause of such high dispersions in 
Section~\ref{sec:envi}.

\section{Discussion} \label{sec:discussion} 

\subsection{Genuine circumnuclear PNe} {\label{sec:iden}

The above spectral analysis confirms that the 77 PN candidates 
selected from the \emph{HST} image are genuine emission line 
objects in the circumnuclear region of M31.  The relationship given
in \cite{Herrmann08} (Figure~\ref{fig:reffcone}) and the BPT diagram (Figure~\ref{fig:reffbpt})
further suggests that the majority of them are consistent with the 
PN nature, rather than compact H~{\sc ii} regions.  Moreover, the 
circumnuclear region of M31 is known to be free of recent or 
on-going star formation (\citealt{Rosenfield12,Dong15}), hence no 
contamination from young, massive stars is expected.  There might 
be, however, other types of emission line objects that could mimic 
the observed spectra of PNe. 

In particular, symbiotic stars are often considered as a major 
source of confusion \citep{Boissay12}.  In our Galaxy, about 1200 
symbiotic star candidates have been found in H$\alpha$ photometric 
survey, however, only $\sim$300 are confirmed to be genuine 
symbiotic stars through spectroscopy \citep{Miszalski13}, which is 
much lower than the number of confirmed Galactic PNe ($\sim$3500, 
\citealp{HASH00}). This may suggest a negligible, if any, 
contamination by symbiotic stars in our sample, unless the actual  
occurrence rate of symbiotic stars in M31's bulge is much different 
from that in our Galaxy, which is unlikely.  Another contamination 
might come from the so-called super-soft X-ray sources (SSSs), 
generally thought to be white dwarfs stably accreting from a 
low-mass companion \citep{Stefano95}.  Ten SSSs have been reported 
in the bulge of M31, based on the {\it Chandra} observations 
\citep{DiStefano06}, but none of them coincides with our HPNe in 
spatial location.  Another source of possible contamination might 
be young, which is compact supernova remnants (SNRs).  However, no 
remnants of core-collapsed SNe are expected in M31's bulge, again 
due to the lack of recent star formation; remnants of Type Ia SNR
might be present, but their number is expected to be small at any 
given time.  Currently only three SNRs are known in the inner bulge 
of M31 \citep{Kong03}.  We thus conclude that the 77 PN candidates 
are most likely genuine PNe. 

\subsection{Environmental effect on the circumnuclear PNe?} \label{sec:envi}

Figure~\ref{fig:sig_sl} indicates that most MMD PNe show value of $\sigma_{5007}$ quite consistent with typical PN velocity dispersions of 10--30${\rm~km~s^{-1}}$ \citep{Kwok00},
whereas a substantial fraction of HPNe exhibit large values of $\sigma_{5007}$ up to $\sim$100${\rm~km~s^{-1}}$.
This difference is unlikely due to different populations of progenitor/central stars. 
Instead, we speculate that environmental effects might have caused the difference. 
In particular, the bulge of M\,31 is known to be filled with a corona of diffuse hot gas \citep{Li07,Li09}.
Materials ejected by stars randomly orbiting in the bulge (or in elliptical galaxies), including stellar winds and PN shells, would inevitably be shock-heated by the hot gas and subsequently become part of the latter, a process fundamental to the cycling and metal-enrichment of the interstellar medium \citep{Mathews90}.
Details of this process were modeled by \cite{Bregman09}, who simulated the interaction of the expanding PN shell with the ambient hot gas of varied temperatures and densities, during which the hot flow transports both heat and momentum to the nebula.  
According to their simulations, part of the nebula can be accelerated to about half of the progenitor star's orbiting velocity (represented by the bulge velocity dispersion).  
Observationally, however, the interaction between a PN and the ambient hot gas has rarely been caught in real time. 
The large $\sigma_{5007}$ seen in some of the HPNe might be providing an indirect evidence for the shock-heated and accelerated shell materials, although this has to be further verified by a self-consistent model suitable for the M\,31 bulge.

\subsection{Dynamical Mass of the inner bulge} \label{sec:mass}

At present, while there has been ample evidence for the existence 
of dark matter (DM) in the outer regions of large disk galaxies 
such as M31, the amount and distribution of DM in the central region 
is poorly known.  In the case of M31, the baryonic mass budget in 
the central $\sim$500~pc region, as reviewed by \cite{Li09}, is 
relatively well determined and overwhelmingly dominated by the 
stars.  The kinematic tracers like PNe has the great potential to 
tightly constrain the mass and even the spatial distribution of DM 
in the central region of a massive galaxy, given a robust 
measurement of the total dynamical mass. 
     
The radial velocities of our HPNe can provide a useful estimate of 
the dynamical mass.  We adopt the projected mass estimator proposed
by \cite{Heisler85}, $GM = f/N{\times}{\sum_{i} V_i\,R_i}$, where 
$V_i$ is the PN radial velocity (from H$\alpha$), $R_i$ is the 
projected galactocentric radius, $G$ is the gravitational constant, 
$N$ is the total number of tracer particles, and $f$=10.2 is a 
scaling parameter depending on the distribution of orbits 
\citep{Heisler85}.  We restrict the analysis on 38 HPNe located 
within 70\arcsec\ ($\sim$267~pc), i.e., the largest radius that 
fits the \emph{HST} field-of-view (the cyan circle in 
Figure~\ref{fig:bulgemap}).  Strictly speaking, this estimator 
should be applied to the tracers of an isotropic velocity 
distribution, which is a reasonably approximation in the inner 
bulge of M31, which again shows only moderate rotation (see the 
analysis in Section~3.3). 

We thus estimate an enclosed mass of 5.9$\pm$0.4$\times$10$^{9}$
$M_{\odot}$ within an effective (de-projected) radius of 340~pc 
(as corrected from the projected 267~pc by a factor of 4/$\pi$). 
The error is estimated using a bootstrapping method.  As explained 
in Section~\ref{sec:kinematics}, this calculation is based on the 
H$\alpha$ velocities.  In order to estimate the systematic error 
introduced by this method, we also use the [O~{\sc iii}] velocities 
and obtained an enclosed mass of 6.9$\pm$0.6$\times$10$^{9}$ 
$M_{\odot}$.  These two masses are consistent with each other 
within the errors. 
     
We contrast the estimate dynamical mass with the latest mass models 
of M31 in the literature.  Combining the SDSS $ugriz$ photometry 
and the \emph{Spitzer} 3.6$\mu$m imaging, \cite{Tamm12} performed 
SED fitting to derive the parameterized 3-dimensional stellar mass 
distribution of M31.  They accounted for the additional baryonic 
components, in particular the H~{\sc i} disk, which provides strong 
constraints on the total gravitational mass at large radii.  This 
in turn allowed them to parameterized the DM distribution, following
either an NFW profile, a Moore profile, or an Einasto profile. 
In the central 340~pc of M31, the total mass is dominated by the 
stellar mass, which, according to the model of \cite{Tamm12}, has 
a value of 3.7$\times$10$^{9}$~$M_{\odot}$; however, a higher value 
of 5.6$\times$10$^{9}$~$M_{\odot}$ was also permitted, as not to 
violate the maximally allowed stellar mass at large radii.  The 
latter value is very close to our estimated dynamical mass, leaving
little room for the DM; while adopting the former value would 
indicate a perhaps unrealistically high amount of the DM.  We note 
that the stellar mass estimate of \cite{Tamm12} was based on a 
Charier initial mass function (IMF), which, as they found, is 
preferred by the mass budget in the disk.  Adopting a steeper IMF 
for the inner bulge could substantially raise the stellar mass. 
We conclude that at present the largest uncertainty in determining 
the DM mass in the inner bulge of M31 probably lies in the 
uncertainty in the stellar IMF.

\section{Summary} \label{sec:summary}

Using the WIYN/Hydra multi-object spectrograph, we have obtained 
fiber spectra for 77 PN candidates in the circumnuclear region of 
M31.  The majority (64\%) of this sample are observed for the first 
time in the spectroscopy.  These spectra are analyzed and results 
discussed in terms of their PN nature, kinematics and dynamics. 
We have also obtained spectra of 300 emission line objects (PNe 
and H~{\sc ii} regions) that were identified in previous surveys; 
most of these objects primarily reside in the M31 disk.  Our main 
results are as follows: 

\begin{itemize}
\item All 77 circumnuclear PN candidates are genuine emission line 
objects, most probably PNe. 

\item Combining our circumnuclear PNe and the disk sample from 
previous observations, the intensity ratio of [N~{\sc 
ii}]$\lambda$6583/H$\alpha$ generally shows a trend of decrease 
with the galactocentric radius, while the [O~{\sc 
iii}]$\lambda$5007/H$\beta$ ratio is generally flat, and large 
scatter exist in the disk sample.

\item The radial velocities of the circumnuclear PNe reveal the 
rotation of the inner bulge, which is consistent with the pattern 
of kinematics about the minor-axis of M31's disk. 

\item From the radial velocities of our sample, we estimate a 
dynamical mass of 6.4$\pm$0.5$\times$10$^{9}$~$M_{\odot}$ for 
the bulge region enclosed within an effective radius of 340~pc. 
\end{itemize}

We emphasize that this is the first effort that combines the 
\emph{HST} narrow-band images and follow-up spectroscopy on a 
ground-based telescope, to identify PN candidates in the inner 
bulge of M31, where the stellar background emission is so strong 
that clean removal of such contamination is extremely difficult. 
In the near future, it is highly desired to extend the spectroscopic
study on the remaining, albeit on average fainter, circumnuclear PN 
candidates in M31.

\acknowledgements
This work is supported by the National Key Research and Development Program of China (No.~2017YFA0402703) and National Science Foundation of China under grant 11133001.
We thank the referee for helpful comments. A.L. and Z.L. are grateful to Eric Hooper and Diane Harmer of the WIYN Observatory for their help with the Hydra observations.
Z.L. acknowledges support from the Recruitment Program of Global Youth Experts. H.D. acknowledges funding support from the European Research Council under the European Union's Seventh Framework
Programme (FP7/2007-2013)/ERC grant agreement n$^{\odot}$[614922]. 

\bibliographystyle{aasjournal}
\bibliography{M31PN_rev.bib}

\clearpage



\end{document}